\documentclass[12pt, reqno]{amsart}
\allowdisplaybreaks

\usepackage{amssymb} 

\newcommand{\morespace}{}

\newcommand{\wt}{\widetilde}
\newcommand{\wh}{\widehat}

\newcommand{\os}[2]{\overset{#1}{#2}}

\newtheorem{thm}{Theorem}

\begin{document}
\morespace

\title[Non-bicolorable configurations]{Non-bicolorable finite configurations of rays and their deformations}
\author[A. Ruuge]{Artur E. Ruuge}
\address{
Department of Quantum Statistics and Field Theory, 
Faculty of Physics, Moscow State University, 
Vorobyovy Gory, 119899, Moscow, Russia 
\emph{and} 
Department of Mathematics and Computer Science, 
University of Antwerp, 
Middelheim Campus Building G, 
Middelheimlaan 1, B-2020, Antwerp, Belgium 
}
\email{Artur.Ruuge@ua.ac.be} 

\begin{abstract}
\morespace
A new infinite family of examples of finite non-bi\-colorable configurations of rays 
in Hilbert space is described.
Such configurations appear in the analysis of quantum mechanics in terms of Bell's inequalities and Kochen-Specker 
theorem and illustrate that there is no measurable space in the background of the probability model of 
a quantum system. 
The mentioned examples are naturally parametrized by a positive integer divisible by four and by several 
complex-valued parameters, whose number depends on this integer. 
In order to compare two configurations with the same number of rays, a notion of deformation of a configuration is 
introduced. The constructed examples are then interpreted as obtained by way of deformations.   
\end{abstract}
\maketitle

\section{Introduction}

The non-bicolorable finite configurations of rays play a substantial role in quantum mechanics. 
Generally speaking, they are meant to illustrate various weird features of quantum theory -- the so called 
contextuality, non-locality, indeterminism,   etc. 
(depending on the metaphysical point of view accepted by a particular scientist). 
This makes them especially interesting, and in fact fundamental, for the quantum computing technology. 

The terminology used in the present paper is close to the one 
introduced in \cite{ZimbaPenrose}. 
Let $\mathcal{H}$ be a Hilbert space over $\mathbb{C}$ of finite dimension $d$. 
Consider the set of all projective lines $\mathbb{P} (\mathcal{H})$, and let $A$ be a subset of it.  
The elements of $A$ are termed as \emph{rays}. Let $\Gamma$ be a finite set consisting of two 
formal symbols termed as \emph{colors}, say $\Gamma := \lbrace \mathtt{red}, \mathtt{blue} \rbrace$.   
Denote by $\perp$ the orthogonality relation on $A$ induced by the inner product on $\mathcal{H}$. 
A function $v : A \to \Gamma$ is called a \emph{bicoloring} if the following two conditions are met: 
1) for every $l, l' \in A$, if 
$v (l')$ and $v (l)$ are both $\mathtt{red}$, then $l \not \perp l'$; 
2) in every collection $l_1, l_2, \dots, l_d \in A$ of $d$ pairwise orthogonal rays, there is an element 
$l_{r}$ such that $v (l_r) = \mathtt{red}$.  
The set $A$ is called \emph{bicolorable} in case it admits a bicoloring, 
and \emph{non-bicolorable} -- otherwise. 

Intuitively, given a collection of rays $A$, it is natural to imagine the possibility of 
performing a ``homotopy process'' over it. 
For example, if $A$ is non-bicolorable, is it possible to vary the positions of rays in such a way that 
it remains non-bicolorable? Is it possible to transform a bicolorable configuration into a non-bicolorable one?   
So, it is better to view the non-bicolorability as a property of the 
\emph{configuration}. 

The physical meaning of non-bicolorability can be illustrated as follows. 
Interpret $\mathcal{H}$ as the Hilbert space associated to some quantum-mechanical system. 
Hence, for every $l \in A$, the corresponding orthogonal projector $\wh{\pi}_{l}$  
represents in $\mathcal{H}$ an observable that may acquire just two values, $0$ and $1$.   
Call this observable $P_l$, and denote $\mathcal{O}_{A} := \lbrace P_l \rbrace_{l \in A}$. 
It may seem natural from the positions of classical physics to think that 
it is possible to construct a non-empty set $\Omega$ and a map 
$\rho : \mathcal{O}_A \to \mathcal{P} (\Omega)$, such that 
$\rho (l')$ and  $\rho (l)$ are disjoint whenever $l \perp l'$, and 
for every collection $l_1, l_2, \dots, l_d$ of pairwise orthogonal rays, 
the sets $\rho (l_i)$, $i = 1, 2, \dots, d$, partition $\Omega$. 
But in this case, one can associate to every point $\omega \in \Omega$ a bicoloring $v_{\omega}$ of $A$: 
$v_{\omega} (l) := \mathtt{red}$, if $\rho (l) \ni \omega$, and $v_{\omega} (l) := \mathtt{blue}$, -- otherwise. 
Hence, for a non-bicolorable $A$ such a map $\rho$ cannot exist. Therefore, the behaviour of the 
system with respect to the observables $P_l$, $l \in A$, will always look non-classical.
For a deeper discussion of the physical meaning of such constructions one may refer to 
\cite{Bell, AccardiFedullo, GHSZ, Meyer, Kent, CliftonKent, BarrettKent, Cabello, Appleby}.
The general motivation can be found in \cite{Bohm, Neumann}.  

It is interesting to mention the link between the notion of non-bicolorability and 
the discussion about indeterminism in quantum physics. 
If one speaks about ``determinism'', one operates with the notions of \emph{cause} and \emph{effect}. 
Whatever happens always has a reason, why it happens. 
A \emph{naive} conception of determinism is based on a set-theoretic understanding of ``causes'' and ``effects''.  
In other words, they are mathematically nothing more but \emph{points} of some sets.   
Invoking the notation above, one is tempted to view the points $\omega \in \Omega$ as causes in the following sense. 
For every \emph{single} measurement act, just one of the points $\omega \in \Omega$ becomes active;  
if $\rho (l) \ni \omega$, then the result of measurement of $P_l$ is destined to be $1$; otherwise it should be $0$. 
Since for a non-bicolorable $A$ such a space $\Omega$ does not exist at all, 
the naively deterministic point of view on the system with respect to observables $\mathcal{O}_A$ must be ruled out. 

A more careful investigation of the physical meaning behind the non-bicolorable configurations 
should take into account the fact that the collection of all observables corresponding to 
the orthogonal projections onto 
$1$-dimensional subspaces is not just a \emph{set}, but a \emph{topological space}
(the topology stems from the inner product on $\mathcal{H}$). 
It is known due to \cite{KochenSpecker}, that 
finite non-bicolorable configurations in 
$\mathcal{H}$ exist
whenever its dimension $d$ satisfies $d \geqslant 3$. 
At the same time, according to \cite{Meyer, Kent, CliftonKent},  
every such space admits a (countable) bicolorable configuration, 
which is \emph{dense} if viewed in the mentioned topology.  
This implies, that if we accept a thesis that no experimental setup 
can acheive an ideal realization of measurements (i.e. one only tries to measure a target observable, 
but the resulting observable being actually measured is not precisely known), 
then one may question the falsifyability of the existence of ``non-contextual hidden variables'' 
in (non-relativistic) quantum mechanics. 
This has been a subject of some non-trivial discussions in recent papers 
(see \cite{BarrettKent, Cabello, Appleby} and references therein). 

It is not difficult to give an example of a non-bicolorable configuration. If the dimension $d$ of the space 
$\mathcal{H}$ satisfies $d \geqslant 3$, then the whole set $\mathbb{P} (\mathcal{H})$ is non-bicolorable. 
This is a straightforward corollary of a classical result in functional analysis -- the Gleason's theorem.   
More important is that there exist \emph{finite} non-bicolorable configurations. 
Kochen and Specker have found \cite{KochenSpecker} the first example of such a configuration for $d = 3$. 
Their construction is quite sophisticated and involves $117$ rays. 
Since then several other examples in spaces of other dimensions have been 
found \cite{ZimbaPenrose, Peres, Mermin, KernaghanPeres, AravindLee-Elkin, CabelloEG, CabelloEG1}. 
It is necessary to note, that they all exhibit some degree of symmetry, and by that 
the non-bicolorability may be viewed as stemming from the properties of the corresponding group. 
Despite of this fact, a complete classification of all finite non-bicolorable configurations is not known and  
so far each time an element of creativity is required to find a new example. 
The aim of the present paper is to describe a new \emph{infinite} 
family (or, more precisely, family of families) of such configurations and to investigate the 
possibility of their deformations.

\section{Non-orthogonality and deformations} 

Let $A$ be a finite collection of rays in a complex Hilbert space $\mathcal{H}$ of finite dimension $d$. 
In order to verify if $A$ is bicolorable or not, it suffices to know just the  
orthogonality relation between its elements, or, what is equivalent, but conceptually better, the 
\emph{non-orthogonality} relation $\not \perp$. 
The formulae defining the rays themselves become at this stage unessential. 
Let us start with the description of this relation for the family of the upcoming examples. 

We need some auxiliary notation first. 
Let $V$ be a \emph{finite} set, $\# V = N$. It will be necessary to assume later that the number 
$4$ divides $N$, 
$N = 4 n$, $n \in \mathbb{N}$, but at this moment it is not important. Now, let $p_0, p_1, p_2, p_3 \in 
\mathbb{Z} / 2$ be four parameters. Look at all functions $\varphi : V \to \mathbb{Z} / 2$ and 
for every $U \subset V$ denote 
\begin{equation} 
L (U) := \big\lbrace \varphi : V \to \mathbb{Z} / 2 \, \big| \, \sum_{v \in U} \varphi (v) = p_{\#_4 U} \big\rbrace, 
\label{index_set_LU}
\end{equation} 
where $\#_4 U \in \mathbb{Z} / 4$ is the cardinality of $U$ modulo $4$. 
Consider the \emph{disjoint} union of the sets $L (U)$. 
Denote it as $X := \bigsqcup_{U \in \mathcal{P} (V)} L (U)$, and let 
$i_U : L (U) \rightarrowtail X$, $U \in \mathcal{P} (V)$, be the canonical injections. 
Define a relation $R$ on $X$ as follows. 
Take any $x, x_1 \in X$ of the form $x = i_U (\varphi)$ and $x_1 = i_{U_1} (\varphi_1)$. 
If $U_1 = U$, then put  
$(x, x_1) \in R :\Leftrightarrow \varphi_1 = \varphi$; 
if $U_1 \not = U$, then put    
\begin{equation} 
(x, x_1) \in R \quad : \Leftrightarrow
\sum_{v \in U \Delta U_1} \varphi (v) = 
\sum_{v \in U \Delta U_1} \varphi_1 (v) + p_{\#_4 (U \Delta U_1)},   
\label{nonorth_rel}
\end{equation}
where $\Delta$ denotes the symmetric difference between the two subsets. 

Observe, that if $U \subset V$ is not empty, then $\# L (U) = 2^{N-1}$. 
Let the dimension $d$ of space $\mathcal{H}$ coincide with this number, $d = 2^{N-1}$. 
Suppose that $A$ can be viewed as a union of $N + 1$ pairwise disjoint subsets
of same cardinality $d$: $N$ sets denoted as  
$A_{v}$, $v \in V$, and a set $\wh{A}$, i.e. $A = ( \cup_{v \in V} A_v ) \cup \wh{A}$. 
Let the elements of $A_v$ be indexed by $L (\lbrace v\rbrace)$, and the elements of 
$\wh{A}$, -- by $L (V)$. Write $A_{v} = \lbrace \Psi_{\sigma}^{v} \rbrace_{\sigma \in L(\lbrace v \rbrace)}$, 
$\wh{A} = \lbrace F_{\pi} \rbrace_{\pi \in L (V)}$. Hence, the set $A$ becomes indexed by a subset $X_0 \subset X$, 
which is a disjoint union of all $L (\lbrace v \rbrace)$, $v \in V$, and $L (V)$. 
Take any $R' \subset R$ such that $\forall U \in \mathcal{P} (V) \, 
\forall \varphi \in L (U): (i_U (\varphi), i_U (\varphi)) \in R'$. 
By this one ensures that $\forall U$ and $\forall \varphi, \varphi'$, 
there is an equivalence $(i_U (\varphi), i_U (\varphi_1)) \in R' \Leftrightarrow 
\varphi = \varphi'$.
At the same time, unlike the case with $R$, for $(i_U (\varphi), i_{U_1} (\varphi_1)) \in R'$, $U_1 \not = U$, 
only an implication ``$\Rightarrow$'' of the form as above in \eqref{nonorth_rel} is valid. 
The relation $R'$ on $X$ induces a relation on $X_0$, which we denote $R_{0}'$. 

Now, suppose that the non-orthogonality relation between the rays that constitute $A$, stems 
precisely from $R_{0}'$. In other words, 
$\Psi_{\sigma}^{v} \not \perp \Psi_{\sigma_1}^{v_1}$ iff  
$(i_{\lbrace v \rbrace} (\sigma), i_{\lbrace v_1 \rbrace} (\sigma_1)) \in R_{0}'$, and 
$\Psi_{\sigma}^{v} \not \perp F_{\pi}$ iff 
$(i_{\lbrace v \rbrace} (\sigma), i_{V} (\pi)) \in R_{0}'$, 
where $v, v_1 \in V$, $\sigma \in L (\lbrace v \rbrace)$, $\sigma_1 \in L (\lbrace v_1 \rbrace)$, 
$\pi \in L (V)$. In particular, this implies that the elements of each $A_v$, $v \in V$, are pairwise orthogonal, 
as well as the elements of $\wh{A}$. 

The requirement, that $A$ is bicolorable yields a condition on the parameters $p_0, p_1, p_2, p_3$. 
Choose and fix a bicoloring, assuming that it exists. In particular, each of the subsets $A_v$, $v \in V$, 
has precisely one $\mathtt{red}$ ray, and the subset $\wh{A}$ has precisely one $\mathtt{red}$ ray. 
Denote the $\mathtt{red}$ rays as $\Psi_{\sigma_v}^{v}$, $v \in V$, and $F_{\wh{\pi}}$. 
Since any two $\mathtt{red}$ rays cannot be orthogonal, taking into 
account the explicit description \eqref{nonorth_rel} of $R$, one obtains: 
\begin{equation*} 
\sum_{z = v, v_1} \big( \sigma_v (z) + \sigma_{v_1} (z) \big) = p_2, \quad
\sum_{z \in V \backslash \lbrace v \rbrace} \big( \sigma_{v} (z) + \wh{\pi} (z) \big) = p_3.  
\end{equation*}
Invoking the definitions \eqref{index_set_LU} of $L (U)$, $U \subset V$, one derives: 
\begin{gather*} 
\sigma_v (v_1) + \sigma_{v_1} (v) = p_2, \\ 
\sum_{z \in V \backslash \lbrace v \rbrace} \sigma_v (z) = \wh{\pi} (v) + p_0 + p_3. 
\end{gather*}
Take a sum over all pairs $\lbrace v, v_1 \rbrace$ in the first formula. 
Similarly, take a sum over all $v$ in the second formula. This yields: 
\begin{gather*} 
\sum_{\substack{v, v_1 \in V, \\ v_1 \not = v}} \sigma_v (v_1) = 
\frac{N (N - 1)}{2} \, p_2, \\ 
\sum_{\substack{v, z \in V, \\ z \not = v}} \sigma_v (z) = 
\sum_{v \in V} \wh{\pi} (v) + N \big( p_0 + p_3 \big). 
\end{gather*} 
Since $\wh{\pi} \in L (V)$, we have $\sum_{v \in V} \wh{\pi} (v) = p_0$.  
Subtracting the first equality from the second one, we obtain: 
\begin{equation*}
(N + 1) p_0 + \frac{N (N - 1)}{2} \, p_2 + N p_3 = 0.
\end{equation*} 
Finally, since $N = 4 n$, $n \in \mathbb{N}$, this simply reduces to $p_0 = 0$. 
Therefore, it is sufficient just to have $p_0 = 1$ in order to claim that the configuration $A$ is non-bicolorable.  

It is natural to consider the following problem. 
Suppose the set $A$ with the described relations between the 
rays exists. Is it possible to vary its configuration without breaking up these relations? 
Consider a simple analogy. Take two orthonormal bases $\lbrace e_i \rbrace_{i = 0}^{3}$ and 
$\lbrace f_j \rbrace_{j = 0}^{3}$ in $\mathbb{C}^4$, and suppose that $\mathbb{C} f_j \not \perp \mathbb{C} e_i$ iff 
$i + j$ is even. The space $\mathbb{C}^4$ splits into orthogonal sum $H_0 \oplus H_1$, with 
$H_0 := \mathrm{span} \lbrace e_0, e_2 \rbrace = \mathrm{span} \lbrace f_0, f_2 \rbrace$ and 
$H_1 := \mathrm{span} \lbrace e_1, e_3 \rbrace = \mathrm{span} \lbrace f_1, f_3 \rbrace$. 
One may rotate infinitesimally the pair $\lbrace f_0, f_2 \rbrace$ in such a way that both of its elements 
are kept in $H_0$. 
At the same time, one may keep the other six vectors fixed, and this will not break up the 
mentioned description of the non-orthogonality relation.  
This motivates the following definition. Let $A$ and $A'$ be two configurations of rays in $\mathcal{H}$, 
$\mathrm{dim} \mathcal{H} = d$. A bijective map $\delta : A \overset{\sim}{\to} A'$ is 
called a \emph{deformation} of $A$, if it satisfies the following condition: 
for all $l, l' \in A$, $\delta (l') \not \perp \delta (l)$ iff $l \not \perp l'$. 
In the next sections we will construct for every $N = 4 n$, $n \in \mathbb{N}$, a non-bicolorable configuration 
with the non-orthogonality relation stemming from $R_{0}'$ as described above, and then prove explicitly that they 
admit non-trivial deformations. By that a new family of families of non-bicolorable configurations is obtained. 

\section{Rays and equations} 

Recall that $N = \# V = 4 n$, $n \in \mathbb{N}$. 
Assign the values to the parameters: $p_0 = 1$, $p_1 = p_2 = p_3 = 0$. 
We need to construct the rays $\Psi_{\sigma}^{v}$, $v \in V$, $\sigma \in L (\lbrace v \rbrace)$, and 
$F_{\pi}$, $\pi \in L (V)$, such that the following implications are valid: 
\begin{gather*} 
\Psi_{\sigma}^{v} \not \perp \Psi_{\sigma_1}^{v_1} \quad \Rightarrow \quad  
\sigma (v_1) + \sigma_1 (v) = 0, \\ 
\Psi_{\sigma}^{v} \not \perp F_{\pi} \quad \Rightarrow \quad 
\sum_{z \in V \backslash \lbrace v \rbrace} (\pi (z) + \sigma (z)) = 0. 
\end{gather*}
Put $\mathcal{H} := \mathbb{C}^2 \otimes \mathbb{C}^2 \otimes \dots \otimes \mathbb{C}^2 $ ($N-1$ times). 
Let us search the $\Psi$-rays in the form $\mathbb{C} f_1 \otimes f_2 \otimes \dots \otimes f_{N-1}$, i.e. 
each of these rays is a one-dimensional subspace spanned over a \emph{homogeneous} vector. 
Having in mind the mentioned property of the $\not \perp$ relation with respect to the $\Psi$-vectors, 
it is convenient to consider a completely connected non-oriented graph with vertices $V$. Assign to the 
edges of this graph the numbers $0, 1, \dots, N - 2$ in such a way, that any two edges that share a common vertex, 
are labeled differently. The most natural way to implement this is the following. 
Identify $V$ with $\mathbb{Z} / (N - 1) \sqcup \lbrace \ast \rbrace$, where $\ast$ is just a formal symbol. 
Write $\mathbb{Z} / (N - 1)$ additively and identify its elements with $0, 1, \dots, N - 2$.  
For every $i, j \in \mathbb{Z} / (N -1)$, $i \not = j$, assign to the edge connecting $i$ and $j$ the number 
$i + j$. To every edge that connects $\ast$ and $k \in \mathbb{Z} / (N - 1)$, the number $2 k$ is assigned. 
By that the required numbering of edges is obtained. Now, choose for every edge an orthonormal basis in $\mathbb{C}^2$, 
and let its elements be indexed by $\mathbb{Z} / 2$. Denote the basis corresponding to the edge 
$\lbrace \ast, k \rbrace$, $k \in \mathbb{Z} / (N - 1)$, as 
$\lbrace \varphi [k]_{\alpha} \rbrace_{\alpha \in \mathbb{Z} / 2}$, and 
the basis corresponding to the edge $\lbrace l, j \rbrace$, $l, j \in \mathbb{Z} / (N - 1)$, $l \not = j$, as 
$\lbrace \psi [\lbrace l, j \rbrace]_{\beta} \rbrace_{\beta \in \mathbb{Z} / 2}$. 
For every $l \in \mathbb{Z} / (N - 1)$, $\sigma \in L (\lbrace l \rbrace)$, $\rho \in L (\lbrace \ast \rbrace)$, 
put 
\begin{gather}
\Psi_{\rho}^{\ast} := \mathbb{C} 
\bigotimes_{k \in \mathbb{Z} / (N - 1)} 
\os{2 k}{\varphi [k]}_{\rho (k)}, 
\label{ray_Psi_star}
\\
\Psi_{\sigma}^{l} := \mathbb{C} \Big\lbrace 
\os{2 l}{\varphi [l]}_{\sigma (\ast)} \otimes 
\Big( 
\bigotimes_{\substack{j \in \mathbb{Z} / (N - 1), \\ j \not = l}} 
\os{l + j}{\psi} [\lbrace l, j \rbrace]_{\sigma (j)}
\Big) 
\Big\rbrace, 
\label{ray_Psi_l}
\end{gather}
where a Feynman-Maslov-type notation is used to determine the place of the factors in a tensor product. 
For instance, an expression of the form $\os{0}{a} \otimes \os{2}{b} \otimes \os{1}{c}$ means nothing more, but 
$a \otimes c \otimes b$. 
Note, that with this notation the symbols under $\otimes$-product commute: for example, 
$\os{0}{a} \otimes \os{2}{b} \otimes \os{1}{c} = 
\os{2}{b} \otimes \os{0}{a} \otimes \os{1}{c}$.  
It is clear, that the required properties of $\not \perp$ relation with respect to $\Psi$-rays are established.  

Now, let us take care about $F_{\pi}$. 
One can write $F_{\pi}$ in the form 
\begin{equation} 
F_{\pi} = \mathbb{C} \sum_{\xi \in L (\lbrace \ast \rbrace)} 
A_{\pi} (\xi) \bigotimes_{k \in \mathbb{Z} / (N - 1)} 
\os{2 k}{\varphi [k]}_{\xi (k)}, 
\label{ray_F}
\end{equation}
where $A_{\pi} (\xi) \in \mathbb{C}$ are some coefficients.  
Note, that $\xi$ and $\pi$ can be completely recovered from their restrictions to $\mathbb{Z} / (N - 1) \subset V$.  
Denote $|\xi| := \sum_{i} \xi (i)$, $|\pi| := \sum_{i} \pi (i)$, $i \in \mathbb{Z} / (N - 1)$. 
The required implication $\Psi_{\xi}^{\ast} \not \perp F_{\pi} \Rightarrow |\pi| + |\xi| = 0$ is equivalent to 
$|\pi| + |\xi| = 1 \Rightarrow \Psi_{\xi}^{\ast} \perp F_{\pi}$. 
Hence, $A_{\pi} (\xi)$ can be non-zero only if $|\pi| + |\xi| = 0$. One has: 
\begin{equation} 
A_{\pi} (\xi) = 0, \quad \text{if $|\pi| + |\xi| = 1$}. 
\label{A_pi_xi_null}
\end{equation} 
Therefore, the collection $\lbrace A_{\pi} (\xi) \rbrace_{\xi, \pi}$ can be viewed as a 
block-diagonal complex matrix with two non-trivial blocks. One of these blocks corresponds to 
$|\pi| = |\xi| = 0$, and the other -- to $|\pi| = |\xi| = 1$. 
Since the rays $\lbrace F_{\pi} \rbrace_{\pi \in L (V)}$ are pairwise orthogonal, 
one has a condition 
\begin{equation} 
\sum_{\xi : |\xi| = p} 
A_{\pi} (\xi)^{*} A_{\pi'} (\xi) \propto \delta_{\pi, \pi'}, 
\label{unitarity_cond}
\end{equation}
where $|\pi| = |\pi'| = p$, $p \in \mathbb{Z} / 2$. 
Let us term \eqref{unitarity_cond} as the `unitarity condition'.

Now consider the implications of the requirements corresponding to the rays $\Psi_{\sigma}^{l}$, 
$l \in \mathbb{Z} / (N - 1)$ and  $F_{\pi}$. If 
$\Psi_{\sigma}^{l} \not \perp F_{\pi}$, then one must have 
$(1 + \pi (l)) + \sigma (\ast) + \sum_{j : j \not = l} \sigma (j) = 0$. 
Note that since the number of elements in $\mathbb{Z} / (N - 1)$ is odd, the multiplication by $2$ is invertible. 
This allows to introduce a variable $k = (l + j)/ 2$ replacing the variable $j$ in the tensor product in the 
formula for $\Psi_{\sigma}^{l}$. One derives: if $\pi (l) + \sigma (\ast) + 
\sum_{j : j \not = l} \sigma (j) = 0$, then 
$F_{\pi} \perp \Psi_{\sigma}^{l}$, i.e. 
\begin{equation} 
\sum_{\xi \in L (\lbrace \ast \rbrace)} A_{\pi} (\xi) \,  
\delta_{\sigma (*), \xi (l)} 
\prod_{\substack{k \in \mathbb{Z} / (N - 1), \\ k \not = l}} 
\big( \psi [\lbrace l, 2 k - l \rbrace]_{\sigma (2 k - l)}, 
\varphi [k]_{\xi (k)} \big) = 0,  
\label{general_A_pi_xi}
\end{equation}
where $(\cdot, \cdot)$ is the inner product in $\mathbb{C}^2$, linear with respect to the second argument. 
For every $l \in \mathbb{Z} / (N - 1)$, this is some collection of equations indexed by $\pi$ and $\sigma$. 
If one first chooses a value for $\pi$, and then for $\sigma$, then one has $2^{N - 1}$ possibilities 
for $\pi$ and $2^{N - 2}$ for $\sigma$. 
In total there are $(N - 1) \times 2^{N - 1} \times 2^{N - 2}$ equations. 
It is necessary to define $\varphi [k]_{\alpha}$ and $\psi [\lbrace l, j \rbrace]_{\beta}$ in such a way, that 
this system has a solution as a linear system with respect to the indeterminates 
$A_{\pi} (\xi)$, $|\pi| = |\xi|$. After that it is necessary to select such a solution 
that satisfies the unitarity conditions. By that one produces a 
non-linear system of equations with respect to the variables $\varphi [k]_{\alpha} \in \mathbb{C}^2$, 
$\alpha \in \mathbb{Z} / 2$, and $\psi [\lbrace l, j \rbrace]_{\beta} \in \mathbb{C}^2$, 
$\beta \in \mathbb{Z} / 2$, which played the role of parameters in the system 
for $A_{\pi} (\xi)$, and now become indeterminates themselves. 
It is not obvious that it has any solution at all. 
Nevertheless, the solutions exist, and in fact there are quite many. 
They are described in the next sections.

\section{Specialization and reduction} 

The main system of equations \eqref{general_A_pi_xi}
in its general form is hardly manageable unless one makes some additional assumptions. 
Denote 
\begin{equation} 
u [i, j]_{\alpha, \beta} := 
\big( \psi [\lbrace i, j \rbrace]_{\beta}, 
\varphi [(i + j)/ 2]_{\alpha} \big), 
\label{matrix_u}
\end{equation}
where $i, j \in \mathbb{Z} / (N -1)$, $i \not = j$; $\alpha, \beta \in \mathbb{Z} / 2$. 
The corresponding $2 \times 2$ matrices $u [i, j]$  are unitary. 
Note that $u [i, j] = u [j, i]$. 
Let us make a crucial assumption, which will let us essentially simplify the equations. 
Specialize $u [i, j]$: 
\begin{equation} 
u [i, j] =  
\frac{1}{\sqrt{2}}
\left( 
\begin{matrix} 
1 & x_{i, j} \\ 
x_{i, j}^{*} & -1
\end{matrix}
\right), \qquad 
|x_{i, j}| = 1. 
\label{special_u}
\end{equation}
It is important that the parameters $x_{i, j}$ have an absolute value $1$. 
Since $u [i, j] = u [j, i]$, one must mention the condition 
\begin{equation*} 
x_{i, j} = x_{j, i}, 
\end{equation*}
where $i$ and $j$ vary over $\mathbb{Z} / (N - 1)$, $i \not = j $. 

The equation \eqref{general_A_pi_xi} for $A_{\pi} (\xi)$ acquires the form:
\begin{equation*} 
\sum_{\substack{\xi \in L (\lbrace \ast \rbrace), \\ \xi (l) = \sigma (\ast)}} A_{\pi} (\xi) \, 
(- 1)^{s_l (\sigma, \xi)} 
\Big( \prod_{\substack{k : k \not = l, \\ \sigma (2 k - l) = 1, \\ \xi (k) = 0}} 
x_{l, 2 k - l} \Big)
\, 
\Big( \prod_{\substack{k : k \not = l, \\ \sigma (2 k - l) = 0, \\ \xi (k) = 1}} 
x_{l, 2 k - l}^{*} \Big)
= 0. 
\end{equation*}
where 
\begin{equation*}
s_l (\sigma, \xi) := \sum_{k : k \not = l} \delta_{\sigma (2 k - l), 1} \delta_{\xi (k), 1}, 
\end{equation*}  
$k$ varies over $\mathbb{Z} / (N - 1)$. 
Multiply the left-hand side of this equation by 
$\prod_{\substack{k : k \not = l, \\ \sigma (2 k - l) = 0}} x_{l, 2 k - l}$. 
Invoking the fact that $|x_{i, j}| \equiv 1$, one obtains: 
\begin{equation*} 
\sum_{\substack{\xi \in L (\lbrace \ast \rbrace), \\ \xi (l) = \sigma (\ast)}} A_{\pi} (\xi) \, 
(- 1)^{s_l (\sigma, \xi)} 
\Big( \prod_{\substack{k : k \not = l, \\ \xi (k) = 0}} 
x_{l, 2 k - l} \Big)
= 0. 
\end{equation*}
Recall, that this kind of equation holds for every $l \in \mathbb{Z} / (N - 1)$, 
every $\pi \in L (V)$, and every $\sigma \in L (\lbrace l \rbrace)$, such that 
$\sigma (\ast) = \pi (l) + \sum_{j : j \not = l} \sigma (j)$. 
Since $A_{\pi} (\xi)$ can be non-zero only if $\xi (l) = |\pi| + \sum_{j: j \not = l} \xi (j)$, 
one may understand the summation in the latter formula as over all $\xi$ that satisfy 
$|\xi| = |\pi|$ and $\sum_{j : j \not = l} \big( \xi (j) + \pi (j) + \sigma (j) \big) = 0$. 
 
Take any $l$, $\pi$, and $\sigma$. 
Observe that if one takes any even number of points of $(\mathbb{Z} / (N - 1)) \backslash \lbrace l \rbrace$, 
and then changes the values of $\sigma$ at these points by adding $1$, 
then the result still satisfies $\sigma (\ast) = \pi (i) + \sum_{j : j \not = i} \sigma (j)$. 
In particular, consider the points $l + 1$ and $l + 2$, and put  
$\sigma_1 (l + 1) = 1 + \sigma (l + 1)$, $\sigma_1 (l + 2) = 1 + \sigma (l + 2)$, and 
$\sigma_1 (v) = \sigma (v)$ for $v \not = l + 1, l + 2$. 
Take the equation corresponding to $(l, \pi, \sigma_1)$, multiply it by $(-1)^{\sigma (l + 1)}$, and add the result 
to the equation corresponding to $(l, \pi, \sigma)$. By this one obtains: 
\begin{equation*} 
\sum_{\substack{\xi \in L (\lbrace \ast \rbrace), \\ \xi (l) = \sigma (\ast)}} A_{\pi} (\xi) \, 
\Lambda_{l}^{(1)} (\sigma, \xi)  
(- 1)^{s_l (\sigma, \xi)} 
\Big( \prod_{\substack{k : k \not = l, \\ \xi (k) = 0}} 
x_{l, 2 k - l} \Big)
= 0, 
\end{equation*} 
where 
\begin{equation*} 
\Lambda_l^{(1)} (\sigma, \xi) := 
1 + (-1)^{\sigma (l + 1)} \, 
(- 1)^{s_l (\sigma_1, \xi) - s_l (\sigma, \xi)}.  
\end{equation*}
Note, that for any integer $q$, one has $(-1)^{q} = (-1)^{-q}$. Since   
\begin{equation*} 
s_l (\sigma_{1}, \xi) + s_l (\sigma, \xi) = 
\sum_{\substack{
k : k \not = l, \\ 
2 k - l \in \lbrace l + 1, l + 2 \rbrace
}} 
(\delta_{\sigma (2 k - l), 1} + \delta_{\sigma (2 k - l), 0}) \delta_{\xi (k), 1}, 
\end{equation*}
the expression for $\Lambda_l^{(1)} (\sigma, \xi)$ reduces to 
\begin{equation*} 
\Lambda_l^{(1)} (\sigma, \xi) = 
1 + (-1)^{\delta_{\sigma (l + 1), 1} + \sum_{s = 1, 2} \delta_{\xi (l + s / 2), 1}}.
\end{equation*}

Next, transform in a similar way the just obtained system of equations for $A_{\pi} (\xi)$ 
(containing $\Lambda_{l}^{(1)} (\sigma, \xi)$)
with respect to the points $(l + 2, l + 3)$, 
after that -- with respect to $(l + 3, l + 4)$, and so forth until $(l + (N-3), l + (N-2))$. 
This yields: 
\begin{equation*} 
\sum_{\substack{\xi \in L (\lbrace \ast \rbrace), \\ \xi (l) = \sigma (\ast)}} A_{\pi} (\xi) \, 
\Big( \prod_{m = 1}^{N - 3} \Lambda_{l}^{(m)} (\sigma, \xi) \Big)   
(- 1)^{s_l (\sigma, \xi)} 
\Big( \prod_{\substack{k : k \not = l, \\ \xi (k) = 0}} 
x_{l, 2 k - l} \Big)
= 0, 
\end{equation*}
where 
\begin{equation*} 
\Lambda_l^{(m)} (\sigma, \xi) = 
1 + (-1)^{\delta_{\sigma (l + m), 1} + \sum_{s = m, m + 1} \delta_{\xi (l + s / 2), 1}}.
\end{equation*}

Now let us compute the quantities $\Lambda_l^{(m)} (\sigma, \xi)$. 
For any $x, y, z \in \mathbb{Z} / 2$, the quantity 
$1 + (-1)^{\delta_{x, 1} + \delta_{y, 1} + \delta_{z, 1}}$ equals $2$, whenever 
$\delta_{x, 1} + \delta_{y, 1} + \delta_{z, 1}$ is even, and $0$, -- otherwise. 
Hence, $1 + (-1)^{\delta_{x, 1} + \delta_{y, 1} + \delta_{z, 1}} = 2 \delta_{x + y + z, 0}$, and one obtains: 
$\Lambda_{l}^{(m)} = 2 \delta_{\sigma (l + m) + \sum_{s = m, m + 1} \xi (l + s / 2), 0}$. 
Therefore, the non-trivial contributions to the sum over $\xi$ in the equation for $A_{\pi} (\xi)$ stem only 
from those terms, which satisfy 
\begin{equation*} 
\sigma (l + m) + \sum_{s = m, m + 1} \xi (l + s / 2) = 0, \qquad m = 1, 2, \dots, N-3. 
\end{equation*}  
Fix $\sigma$ and $l$, and put $\xi (l + 1 / 2) = q$, $q \in \mathbb{Z} / 2$. The rest of the values 
$\xi (l + (m + 1) / 2)$, $m = 2, 3, \dots, N - 2$, become determined. 
Since $q$ can be only $0$ or $1$, there are at most two non-trivial terms in the sum. 

It is more convenient to index the equations not by $\sigma$, but by the non-trivial terms themselves. 
Take any equation and take any of its non-trivial terms. Denote the corresponding value of $\xi$ by $\xi_0$. 
The corresponding value $\sigma_0$ of $\sigma$ is then recovered as follows. 
For $m = 1, 2, \dots, N - 3$, one has $\sigma_0 (l + m) = \xi_0 (l + m/2) + \xi_0 (l + (m + 1)/2)$. 
The remaining $\sigma_0 (l + (N - 2))$ is determined by 
$\sum_{j : j \not = l} \big( \xi_0 (j) + \pi (j) + \sigma_0 (j) \big) = 0$. 

Consider the equation corresponding to $(l, \pi, \sigma_0)$. 
The second non-trivial term of the sum over $\xi$ corresponds to $\xi = \xi_{0}^{l}$, determined by 
$\xi_{0}^{l} (j) = 1 + \xi_0 (j)$, $j \in \mathbb{Z} / (N - 1)$, $j \not = l$, and 
$|\xi_{0}^{l}| = |\pi|$. The latter implies 
$\xi_{0}^{l} (l) = \sum_{j : j \not = l} (1 + \xi_0 (j)) + |\pi|$. Since $N$ is even, the number of terms in 
the latter sum over $j$, which is $N - 2$, is also even. 
Taking into account, that $|\xi_0| = |\pi|$, one obtains: 
$\xi_{0}^{l} (l) = \xi_0 (l)$. The equation acquires the form: 
\begin{equation*} 
 X_{0}^{(l)} (\xi_{0}) \, 
(-1)^{s_l (\sigma_0, \xi_0)} \, 
A_{\pi} (\xi_0) + 
 X_{1}^{(l)} (\xi_{0}) \, 
(-1)^{s_l (\sigma_0, \xi_0^l)} \, 
A_{\pi} (\xi_0^l) = 0. 
\end{equation*}
where $X_{p}^{(l)} (\xi_{0}) :=  
\prod_{\substack{k : k \not = l, \\ \xi_0 (k) = p}} x_{l, 2 k - l}$.
We need to compute $(-1)^q$, $q := s_l (\sigma_0, \xi_0) + s_l (\sigma_0, \xi_0^l)$, 
i.e. to find out when $q$ is even and when it's odd. 
Substituting the expressions for $\xi_0^l$, one obtains 
$q = \sum_{k : k \not = l} \delta_{\sigma_0 (2 k - l), 1}$. Hence, $q$ is even iff 
$\sum_{m = 1}^{N - 2} \sigma_0 (l + m) = 0$. But this sum is just 
$\sum_{j : j \not = l} (\xi_0 (j) + \pi (j))$. Invoking the fact $|\pi| = |\xi_0|$, one finally obtains 
$q = \xi_0 (l) + \pi (l)$. The equation, after dropping down the index $0$ near $\xi_0$ and $\xi_0^l$, becomes 
\begin{equation} 
\Big( 
\prod_{\substack{k : k \not = l, \\ \xi (k) = 0}} 
x_{l, 2 k - l}
\Big) \, 
A_{\pi} (\xi) +    
(-1)^{\xi (l) + \pi (l)} \, 
\Big( 
\prod_{\substack{k : k \not = l, \\ \xi (k) = 1}} 
x_{l, 2 k - l}
\Big) \, 
A_{\pi} (\xi^l) = 0. 
\label{reduced_A_pi_xi}
\end{equation}
Recall, that the parameter $l$ varies over $\mathbb{Z} / (N - 1)$, 
$\pi \in L (V)$, and $\xi$ is any element of $L (\lbrace \ast \rbrace)$, such that 
$\sum_{i \in \mathbb{Z} / (N - 1)} (\xi (i) + \pi (i)) = 0$; 
$\xi^l \in L (\lbrace \ast \rbrace)$ is defined by $\xi^l (j) := 1 + \xi (j)$, $j \not = l$, and 
$\xi^l (l) = \xi (l)$ (of course, $\xi^l (\ast) = 0$).   
It is  not difficult to solve this system with respect to $A_{\pi} (\xi)$ and obtain a condition on 
$\lbrace x_{i, j} \rbrace_{i, j}$ as a condition of solvability. 
This is implemented in the next sections. 

\section{Divisibility by four}

Recall, that we have made an assumption about the number of points $N$ in 
$V = (\mathbb{Z} / (N - 1)) \sqcup \lbrace \ast \rbrace$. It has to be not just even, but divisible 
by $4$, i.e. $N = 4 n$, $n \in \mathbb{N}$. 
Let us find some solutions for the reduced system of equations \eqref{reduced_A_pi_xi} for $A_{\pi} (\xi)$ and 
illustrate how the mentioned assumption emerges. 

We have not obtained yet a condition on $\lbrace x_{i, j} \rbrace_{i, j}$ that ensures solvability of the system. 
Recall that $|x_{i, j}| \equiv 1$. Nevertheless, let us try if it's possible to put all $x_{i, j} = 1$. 
In this case one has: 
\begin{equation*} 
A_{\pi} (\xi) + (-1)^{\xi (l) + \pi (l)} A_{\pi} (\xi^l) = 0, 
\end{equation*}  
for every $l$ and $\xi$, $\pi$, such that $|\xi| = |\pi|$. 
Rewrite $(-1)^{\xi (l) + \pi (l)}$ as $(-1)^{\delta_{\xi (l), 1 + \pi (l)}}$. 
Note, that since $|\xi| = |\pi|$, and since $N - 1$ is odd, the number of points in 
$\lbrace i \in \mathbb{Z} / (N - 1) | \xi (i) \not = \pi (i) \rbrace$ is even. 
Hence, for $|\xi| = |\pi|$, one may define 
\begin{equation} 
a_{\pi} (\xi) := (-1)^{\frac{1}{2} \sum_{i} \delta_{\xi (i), 1 + \pi (i)}}. 
\label{small_a_pi_xi}
\end{equation} 
Put $a_{\pi} (\xi) := 0$ in case $|\xi| = 1 + |\pi|$. 
Claim, that $A_{\pi} (\xi) = a_{\pi} (\xi)$ is a solution of the considered system. 
Indeed, one needs to show, that 
\begin{equation*} 
(-1)^{q_1 / 2} + (-1)^{\delta_{\xi (l), 1 + \pi (l)}} (-1)^{q_2 / 2} = 0,  
\end{equation*}
where $q_1 := \sum_{i} \delta_{\xi (i), 1 + \pi (i)}$, and 
$q_2 := \sum_{i} \delta_{\xi^l (i), 1 + \pi (i)}$. 
But this is equivalent to $1 + (-1)^q = 0$, where 
\begin{equation*} 
q := \delta_{\xi (l), 1 + \pi (l)} + 
\frac{1}{2} \Big\lbrace 
2 \delta_{\xi (l), 1 + \pi (l)} + 
\sum_{i : i \not = l} 
\Big[ 
\delta_{\xi (i), 1 + \pi (i)} + 
\delta_{1 + \xi (i), 1 + \pi (i)} 
\Big]
\Big\rbrace. 
\end{equation*}
Therefore, one arrives at  
\begin{equation*}
1 + (-1)^{\frac{N - 2}{2}} = 0. 
\end{equation*}
The latter is true for $N = 4n$, and false for $N = 4n + 2$.  

We have just found a solution of the system \eqref{reduced_A_pi_xi} for $A_{\pi} (\xi)$ 
in case all $x_{i, j} = 1$. 
Now let us prove, that the unitarity condition is fulfilled as well. 
Take any $r \in \mathbb{Z} / 2$ and take any $\pi$ and $\pi'$, such that $|\pi| = |\pi'| = r$. 
Claim: 
\begin{equation*} 
\sum_{\xi : |\xi| = r} 
a_{\pi}^{*} (\xi) a_{\pi'} (\xi) = 2^{N - 2} \, \delta_{\pi, \pi'}. 
\end{equation*}

Let us first consider the case $\pi = \pi'$. 
It is necessary to show that 
$\sum_{\xi : |\xi| = r} (-1)^{\sum_{i} \delta_{\xi (i), 1 + \pi (i)}} = 2^{N -2}$.  
Consider the identity: 
\begin{equation*} 
\sum_{i} \big( \delta_{\xi (i), 1} - \delta_{\pi (i), 1} \big) \equiv
\sum_{i} \Big[  
\delta_{\xi (i), 1} \delta_{\pi (i), 0} - 
\delta_{\xi (i), 0} \delta_{\pi (i), 1} 
\Big]. 
\end{equation*}
One needs to compute the sum 
\begin{equation*}
\sum_{i} \delta_{\xi (i), 1 + \pi (i)} = 
\sum_{i} \Big[  
\delta_{\xi (i), 1} \delta_{\pi (i), 0} +  
\delta_{\xi (i), 0} \delta_{\pi (i), 1} 
\Big].
\end{equation*}
Taking into account that the signs of the terms of the sum when it stands in the exponent of $-1$ are 
unessential, one can apply the identity above and derive: 
\begin{equation*} 
(-1)^{\sum_{i} \delta_{\xi (i), 1 + \pi (i)}} = 
(-1)^{\sum_{i} ( \delta_{\xi (i), 1} - \delta_{\pi (i), 1} )} = 
(-1)^{|\xi| + |\pi|} = 1.  
\end{equation*}
Taking the sum over $\xi$, $|\xi| = r$, one obtains $2^{N - 2}$, just as required. 

Now consider the case $\pi \not = \pi'$. 
Recall that $|\pi| = |\pi'| = r \in \mathbb{Z} / 2$. 
It is necessary to show that  
$\sum_{\xi : |\xi| = r} (-1)^{\varkappa (\xi, \pi, \pi') / 2} = 0$, where 
\begin{equation*} 
\varkappa (\xi, \pi, \pi') := 
\sum_{k \in \mathbb{Z} / (N - 1)} \big( 
\delta_{\xi (k), 1 + \pi (k)} + \delta_{\xi (k), 1 + \pi' (k)}
\big). 
\end{equation*}  
Consider a sum $S := \sum_{k} \delta_{\xi (k), 1} [ \delta_{\pi (k), 1} + \delta_{\pi' (k), 1} ]$, 
and rewrite it in two different ways: 
\begin{gather*}
S  = 2 \sum_{k} \delta_{\xi (k), 1} - 
\sum_{k} \delta_{\xi (k), 1} \big[ 
\delta_{\pi (k), 0} + \delta_{\pi' (k), 0}
\big], \\
S = \sum_{k} \big[ 
\delta_{\pi (k), 1} + \delta_{\pi' (k), 1}
\big] - 
\sum_{k} \delta_{\xi (k), 0} 
\big[ 
\delta_{\pi (k), 1} + \delta_{\pi' (k), 1}
\big]. 
\end{gather*}
Subtracting one equality from the other and regrouping the terms, one obtains an identity 
\begin{multline*} 
- 2 \sum_{k} \delta_{\xi (k), 1} 
+ \sum_{k} \big[ \delta_{\pi (k), 1} + \delta_{\pi' (k), 1} \big] = \\ = 
- \sum_{k} \delta_{\xi (k), 1} \big[ 
\delta_{\pi (k), 0} + \delta_{\pi' (k), 0}
\big]
+ \sum_{k} \delta_{\xi (k), 0} 
\big[ 
\delta_{\pi (k), 1} + \delta_{\pi' (k), 1}
\big]. 
\end{multline*}
Use this fact to transform $\varkappa (\xi, \pi, \pi')$: 
\begin{multline*} 
\varkappa (\xi, \pi, \pi') = 
\sum_{k} \sum_{s = 0, 1} \delta_{\xi (k), s} \big( 
\delta_{1 + \pi (k), s} + \delta_{1 + \pi' (k), s}
\big) = \\ = 
2 \sum_{k} \delta_{\xi (k), 1} \big[ 
\delta_{\pi (k), 0} + \delta_{\pi' (k), 0} 
\big] - 2 \sum_{k} \delta_{\xi (k), 1} + 
\sum_{k} \big[ 
\delta_{\pi (k), 1} + \delta_{\pi' (k), 1}
\big]. 
\end{multline*}
Taking into account that $(-1)^{\sum_{k} \delta_{\xi (k), 1}} = (-1)^{|\xi|} = (-1)^r$, 
one reduces the problem of the proof of $\sum_{\xi : |\xi| = r} (-1)^{\varkappa (\xi, \pi, \pi') / 2} = 0$ to 
\begin{equation*} 
\sum_{\xi : |\xi| = r} 
(-1)^{\sum_{k} \delta_{\xi (k), 1} \big[ 
\delta_{\pi (k), 0} + \delta_{\pi' (k), 0} 
\big]} = 0. 
\end{equation*}
Note, that since $\delta_{\pi (k), 0} + \delta_{\pi' (k), 0}$ is $0$ or $2$ if $\pi (k) = \pi' (k)$, 
and $1$ -- otherwise,  the sum over $k$ in the latter formula 
becomes the sum of $\delta_{\xi (k), 1}$ over all $k$, such that $\pi (k) \not = \pi' (k)$.  
Now look at the sets: 
\begin{gather*} 
\lbrace{k | \pi (k) = 0} \rbrace = 
\bigsqcup_{p \in \mathbb{Z} / 2} 
\lbrace{k | \pi (k) = 0 \,\&\, \pi' (k) = p} \rbrace, \\ 
\lbrace{k | \pi' (k) = 0} \rbrace = 
\bigsqcup_{p \in \mathbb{Z}/ 2} \lbrace{k | \pi (k) = p \,\&\, \pi' (k) = 0} \rbrace. 
\end{gather*}
The cardinalities of both sets in the left-hand sides viewed in $\mathbb{Z} / 2$ are $1 + r$. 
Apply $\# (\cdot)$ to the left and right-hand sides and subtract the two equalities. 
Due to the mentioned fact, the number 
\begin{equation*} 
m := \# \lbrace k \, | \, \pi (k) \not = \pi' (k) \rbrace
\end{equation*} 
is even. Note, that $m \not = N - 1$, since $N - 1$ is odd, and $m \not = 0$, since $\pi \not = \pi'$. 
For every $\xi$, $|\xi| = r$, one may consider 
\begin{gather*} 
p := \# \lbrace k \,|\, \xi (k) = 1 \,\&\, \pi (k) \not = \pi' (k) \rbrace, \\
q := \# \lbrace k \,|\, \xi (k) = 1 \,\&\, \pi (k) = \pi' (k) \rbrace. 
\end{gather*}
Since $p + q$ is just the cardinality of $\lbrace k \, | \, \xi (k) = 1 \rbrace$, one obtains 
$[p + q]_2 = r$, where $[\cdot]_2 : \mathbb{Z} \twoheadrightarrow \mathbb{Z} / 2$ is the canonical epimorphism. 
For any given $p$ and $q$, $0 \leqslant p \leqslant m$, $0 \leqslant q \leqslant N - 1 - m$, $[p + q]_2 = r$, 
there exist precisely $C_{m}^{p} C_{N- 1 - m}^{q}$ ways to choose the corresponding $\xi$ ( 
$C_{m}^{p}$ and $C_{N- 1 - m}^{q}$ are the binomial coefficients).  
Hence, the required sum becomes:  
\begin{equation*} 
\sum_{\xi : |\xi| = r} (-1)^{\sum_{k : \pi (k) \not = \pi' (k)} \delta_{\xi (k), 1}} = 
\sum_{p = 0}^{m} \sum_{\substack{q = 0, \\ [p + q]_2 = r}}^{N - 1 - m} 
C_{m}^{p} C_{N - 1 - m}^{q} (-1)^p.  
\end{equation*}
All that remains to show is that the expression in the right-hand side vanishes. 
Denote it by $S$ and rewrite as follows: 
\begin{equation*} 
S = \bigg( 
\sum_{\substack{p = 0, \\ [p]_2 = r}}^{m} 
\sum_{\substack{q = 0, \\ [q]_2 = 0}}^{N - 1 - m} + 
\sum_{\substack{p = 0, \\ [p]_2 = 1 + r}}^{m} 
\sum_{\substack{q = 0, \\ [q]_2 = 1}}^{N - 1 - m}  
\bigg) C_{m}^{p} C_{N - 1 - m}^{q} (-1)^p.  
\end{equation*}
Observe, that 
\begin{equation*} 
\sum_{\substack{q = 0, \\ [q]_2 = 0}}^{N - 1 - m} 
C_{N - 1 - m}^{q} = 
\sum_{\substack{q = 0, \\ [q]_2 = 1}}^{N - 1 - m}  
C_{N - 1 - m}^{q},  
\end{equation*}
since the difference between these two expressions is equal to $(1 + (-1))^{N - 1 - m} = 0$ 
(recall, that $m \not = N - 1$). 
Therefore, $S$ can be written as  
\begin{equation*} 
S = \bigg( \sum_{\substack{q = 0, \\ [q]_2 = 0}}^{N - 1 - m} 
C_{N - 1 - m}^{q} \bigg) 
\sum_{p = 0}^{m} C_{m}^{p} (-1)^{p}. 
\end{equation*}
But the sum over $p$ is just $(1 + (-1))^{m} = 0$ (recall, that $m \not = 0$). 
Hence $S = 0$, and this completes the proof of the unitarity for $a_{\pi} (\xi)$. 

Note that the fact that $4$ divides $N$ has been used only in the proof that $a_{\pi} (\xi)$ 
is a solution of the system of equations for $A_{\pi} (\xi)$. 
The fact that $\lbrace a_{\pi} (\xi) \rbrace_{\xi, \pi}$ satisfy the 
unitarity condition relies just on the assumption that $N$ is even.

\section{General case}

Note that one can rewrite the system of equations \eqref{reduced_A_pi_xi} for $A_{\pi} (\xi)$ as follows: 
\begin {equation*}
\Big( 
\prod_{\substack{k : k \not = l, \\ \xi_{\pi} (k) = 0}} 
x_{l, 2 k - l}^{1 - 2 \delta_{\pi (k), 1}}
\Big) \, 
A_{\pi} (\xi) +    
(-1)^{\xi_{\pi} (l)} \, 
\Big( 
\prod_{\substack{k : k \not = l, \\ \xi_{\pi} (k) = 1}} 
x_{l, 2 k - l}^{1 - 2 \delta_{\pi (k), 1}}
\Big) \, 
A_{\pi} (\xi^l) = 0, 
\end{equation*}
where $\xi_{\pi} (i) := \xi (i) + \pi (i)$, $i \in \mathbb{Z} / (N - 1)$. 
Therefore it suffices to investigate only the subsystem corresponding to, say, $\pi (i) \equiv 0$. 
In order to obtain a solution for a general $\pi$, one needs to replace 
each $x_{i, j}$ with $x_{i, j}^{1 - 2 \delta_{\pi ((i + j) / 2), 1}}$, and each $\xi (i)$ with $\xi_{\pi} (i)$, i.e. 
\begin{equation} 
A_{\pi} (\xi) = C_{\pi} A (\xi_{\pi}), \qquad x_{i, j} \to x_{i, j}^{1 - 2 \delta_{\pi ((i + j) / 2), 1}},  
\label{A_xi}
\end{equation}  
where $C_{\pi} \in \mathbb{C}$, and 
$A (\cdot)$ is the $A_{\pi} (\cdot)$ corresponding to $\pi (i) \equiv 0$.   
Since the coefficients $A_{\pi} (\xi)$ in \eqref{ray_F} are meant to define the rays $F_{\pi}$, 
it is necessary to assume $C_{\pi} \not = 0$. 
Without loss of generality, $C_{\pi} = 1$.  

Put all $\pi (i) = 0$. Let us derive an expression for $A (\xi)$.  
Note, that for $|\xi| = 1$, the equation \eqref{A_pi_xi_null} implies $A (\xi) = 0$. 
Take any $\xi$, $|\xi| = 0$. 
Since $N - 1$ is odd, this implies that $\xi (i)$ takes a value $0$ in odd number of 
points $i \in \mathbb{Z} / (N - 1)$, and 
$1$ in even number of points. 
Consider the set $\alpha$ of points of $\mathbb{Z} / (N - 1)$ where $\xi$ has the value $0$, 
and equip it with a numbering of the form   
$\alpha = \lbrace \alpha_{0}, \alpha_{\pm 1}, \dots, \alpha_{\pm s} \rbrace$ (in total there are $2 s + 1$ points). 
Similarly, consider the set of points of $\mathbb{Z} / (N - 1)$ where the value of $\xi$ is $1$, and 
equip it with a numbering of the form     
$\beta = \lbrace \beta_{\pm 1}, \dots, \beta_{\pm q} \rbrace$ (in total there are $2 q$ points). 
One has: $(2 s + 1) + 2 q = N - 1$.  
One may assume that $\alpha_{i} < \alpha_{i'}$ iff $i < i'$, and 
$\beta_{j} < \beta_{j'}$ iff $j < j'$.  
From the equation for $A (\xi)$, corresponding to $l = \alpha_{s}$, one obtains: 
\begin{equation*} 
A \bigg( 
\begin{matrix} 
\text{$0$ at $\alpha$,}\\
\text{$1$ at $\beta$}
\end{matrix}
\bigg) = - \frac{ 
\prod_{k \in \beta} x_{\alpha_s, 2 k - \alpha_s}
}{ 
\prod_{k \in \alpha \backslash \lbrace \alpha_s \rbrace} 
x_{\alpha_s, 2 k - \alpha_s} 
} \,  
A \bigg( 
\begin{matrix} 
\text{$0$ at $\beta \cup \lbrace \alpha_s \rbrace$,}\\
\text{$1$ at $\alpha \backslash \lbrace \alpha_s \rbrace$}
\end{matrix}
\bigg). 
\end{equation*}
Similarly, from the equation for $A (\text{$0$ at $\beta \cup \lbrace \alpha_s \rbrace$}, 
\text{$1$ at $\alpha \backslash \lbrace \alpha_s \rbrace$})$, corresponding to 
$l = \alpha_{- s}$, one obtains: 
\begin{equation*} 
A \bigg( 
\begin{matrix} 
\text{$0$ at $\beta \cup \lbrace \alpha_s \rbrace$,}\\
\text{$1$ at $\alpha \backslash \lbrace \alpha_s \rbrace$}
\end{matrix}
\bigg) = 
\frac{
\prod_{k \in \alpha \backslash \lbrace \alpha_{\pm s} \rbrace} 
x_{\alpha_{-s}, 2 k - \alpha_{-s}}
}{
\prod_{k \in \beta \cup \lbrace \alpha_{-s} \rbrace} 
x_{\alpha_{-s}, 2 k - \alpha_{-s}}
} \, 
A \bigg( 
\begin{matrix} 
\text{$0$ at $\alpha \backslash \lbrace \alpha_{\pm s} \rbrace$},\\
\text{$1$ at $ \beta \cup \lbrace \alpha_{\pm s} \rbrace$}
\end{matrix}
\bigg), 
\end{equation*}
(this time there is no minus sign before the fraction). 
It is convenient to denote $y (k, j) := x_{j, 2 k - j}$, and put formally $x_{i, i} \equiv 1$. Hence $y (i, i) \equiv 1$.  
Substituting the latter formula into the formula before it, and using the mentioned 
formal notation, one obtains: 
\begin{multline*} 
A \bigg( 
\begin{matrix} 
\text{$0$ at $\alpha$,}\\
\text{$1$ at $\beta$}
\end{matrix}
\bigg) = 
\bigg\lbrace 
\bigg( 
\prod_{k \in \alpha} \frac{y (k, \alpha_{-s})}{y (k, \alpha_{s})}
\bigg) \, 
\bigg( 
\prod_{k \in \beta} \frac{y (k, \alpha_{s})}{y (k, \alpha_{-s})}
\bigg)
\bigg\rbrace 
\times \\ \times 
\frac{-1}{y^2 (\alpha_{s}, \alpha_{-s})}
A \bigg( 
\begin{matrix} 
\text{$0$ at $\alpha \backslash \lbrace \alpha_{\pm s} \rbrace$},\\
\text{$1$ at $ \beta \cup \lbrace \alpha_{\pm s} \rbrace$}
\end{matrix}
\bigg). 
\end{multline*}
Now, perform the same trick, but this time using $\alpha_{\pm (s - 1)}$, 
with the $A (\cdot)$ variable standing in the right-hand side of the latter formula. 
This yields  
\begin{multline*} 
A \bigg( 
\begin{matrix} 
\text{$0$ at $\alpha \backslash \lbrace \alpha_{\pm s} \rbrace$},\\
\text{$1$ at $ \beta \cup \lbrace \alpha_{\pm s} \rbrace$}
\end{matrix}
\bigg) 
= \\ = 
\bigg\lbrace 
\bigg( 
\prod_{k \in \alpha \backslash \lbrace \alpha_{\pm s} \rbrace} 
\frac{y (k, \alpha_{- (s - 1)})}{y (k, \alpha_{s - 1})}
\bigg) \, 
\bigg( 
\prod_{k \in \beta \cup \lbrace \alpha_{\pm s} \rbrace} 
\frac{y (k, \alpha_{s - 1})}{y (k, \alpha_{- (s - 1)})}
\bigg)
\bigg\rbrace 
\times \\ \times 
\frac{-1}{y^2 (\alpha_{s - 1}, \alpha_{- (s - 1)})}
A \bigg( 
\begin{matrix} 
\text{$0$ at $\alpha \backslash \lbrace \alpha_{\pm s}, \alpha_{\pm (s - 1)} \rbrace$},\\
\text{$1$ at $ \beta \cup \lbrace \alpha_{\pm s}, \alpha_{\pm (s - 1)} \rbrace$}
\end{matrix}
\bigg). 
\end{multline*}
Substitute this result into the formula above and rewrite the expression 
so that there appear again the products of the form $\prod_{k \in \alpha}$ and $\prod_{k \in \beta}$. 
One obtains: 
\begin{multline*} 
A \bigg( 
\begin{matrix} 
\text{$0$ at $\alpha$,}\\
\text{$1$ at $\beta$}
\end{matrix}
\bigg) = 
 \bigg( 
\prod_{k \in \lbrace \alpha_{\pm s} \rbrace}  
\frac{y^2 (k, \alpha_{s-1})}{y^2 (k, \alpha_{- (s - 1)})}
\bigg) 
\times \\ \times 
\bigg\lbrace 
\prod_{m = s, s - 1}
\bigg( 
\prod_{k \in \alpha} \frac{y (k, \alpha_{-m})}{y (k, \alpha_{m})}
\bigg) \, 
\bigg( 
\prod_{k \in \beta} \frac{y (k, \alpha_{m})}{y (k, \alpha_{-m})}
\bigg)
\bigg\rbrace 
\times \\ \times 
\bigg[ 
\prod_{m = s, s - 1}
\frac{-1}{y^2 (\alpha_{m}, \alpha_{- m})}
\bigg] \, 
A \bigg( 
\begin{matrix} 
\text{$0$ at $\alpha \backslash \lbrace \alpha_{\pm s}, \alpha_{\pm (s - 1)} \rbrace$},\\
\text{$1$ at $ \beta \cup \lbrace \alpha_{\pm s}, \alpha_{\pm (s - 1)} \rbrace$}
\end{matrix}
\bigg). 
\end{multline*}
Proceeding this way, one finally arrives at a formula that has in its right-hand side a variable of the form 
$A (0_{\alpha_0})$, where $0_{\alpha_0}$ denotes a function $V \to \mathbb{Z} / 2$ 
the restriction of which to $\mathbb{Z} / (N - 1)$ 
is $0$ only at one point $\alpha_0$, and $1$ -- otherwise.  
The result is of the form: 
\begin{multline} 
A \bigg( 
\begin{matrix} 
\text{$0$ at $\alpha$,}\\
\text{$1$ at $\beta$}
\end{matrix}
\bigg) = 
\bigg( 
\prod_{1 \leqslant m < n \leqslant s}
\frac{y^2 (\alpha_{n}, \alpha_m) y^2 (\alpha_{-n}, \alpha_{m})
}{
y^2 (\alpha_{n}, \alpha_{-m}) y^2 (\alpha_{-n}, \alpha_{-m})}
\bigg) 
\times \\ \times 
\bigg\lbrace 
\prod_{m = 1}^{s} 
\bigg( 
\prod_{k \in \alpha} \frac{y (k, \alpha_{-m})}{y (k, \alpha_{m})}
\bigg) \, 
\bigg( 
\prod_{k \in \beta} \frac{y (k, \alpha_{m})}{y (k, \alpha_{-m})}
\bigg)
\bigg\rbrace 
\times \\ \times 
\bigg[ 
\prod_{m = 1}^{s}
\frac{-1}{y^2 (\alpha_{m}, \alpha_{- m})}
\bigg] \, A (0_{\alpha_0}). 
\label{A_al_be}
\end{multline}
Recall, that $y (k, j) := x_{j, 2 k - j}$, $k \not = j$, and $y (i, i) := 1$. 
It remains to express $A (0_{\alpha_0})$ from the corresponding equation with $l = \alpha_0$. One has: 
\begin{equation} 
A (0_{\alpha_0}) = - \bigg( 
\prod_{k = 0}^{N - 2} y (k, \alpha_0)
\bigg) \, A (\mathbf{0}), 
\label{A_zero_al}
\end{equation}
where $\mathbf{0}$ is 
a function $V \to \mathbb{Z} / 2$, the restriction of which to $\mathbb{Z} / (N - 1)$ is 
a constant function with value $0$. 
Hence, all $A (\xi)$, except $A (\mathbf{0})$, are determined. The value of $A (\mathbf{0})$ 
remains arbitrary. 
Since $A (\xi) \not \equiv 0$ is required, 
without loss of generality, one may put $A (\mathbf{0}) = 1$.

Let us compare these formulae with the result of the previous section. Recall, that in order to obtain 
$A_{\pi} (\xi)$, $|\xi| = |\pi|$, it is necessary to 
to take an expression for $A (\xi_{\pi})$  and replace the parameters $x_{i, j}$ with 
$x_{i, j}^{1 - 2 \delta_{\pi ((i + j) / 2), 1}}$. 
The latter is equivalent to replacing $y (k, j)$ with $y (k, j)^{1 - 2 \delta_{\pi (k), 1}}$. 
If one specializes all $y (k, j)$ to $1$, one obtains $A_{\pi} (\xi) = c_{\pi} \, (-1)^{s_{\pi} + 1}$, where 
$s_{\pi} := (\# \lbrace i \, | \, \xi_{\pi} (i) = 0 \rbrace - 1)/ 2$, and $c_{\pi}$ is an arbitrary constant. 
One can also define $q_{\pi} := \# \lbrace i \, | \, \xi_{\pi} (i) = 1 \rbrace / 2$. Since 
$2 q_{\pi} + (2 s_{\pi} + 1) = N - 1$, one has $(-1)^{s_{\pi} + 1} = (-1)^{(N - 2)/2 - q_{\pi} + 1}$. 
Recalling that $4$ divides $N$, one derives that $(N - 2)/ 2$ is odd, 
and therefore $(-1)^{s_{\pi} + 1} = (-1)^{q_{\pi}}$. 
But $q_{\pi}$ is precisely $(1/ 2) \sum_{i} \delta_{\xi (i), 1 + \pi (i)}$, i.e. if one puts $c_{\pi} = 1$, then 
$A_{\pi} (\xi)$ becomes $a_{\pi} (\xi)$, -- the solution \eqref{small_a_pi_xi} described in the previous section. 

Consider any solution of the system for $A_{\pi} (\xi)$ and separate the factor $a_{\pi} (\xi)$, i.e. 
write $A_{\pi} (\xi) = a_{\pi} (\xi) \wt{A}_{\pi} (\xi)$.  
The substitution of the latter expression into the equation \eqref{reduced_A_pi_xi} 
for $A_{\pi} (\xi)$ corresponding to $(l, \pi, \xi)$ 
yields a similar equation for $\wt{A}_{\pi} (\xi)$, which looks almost the same and has just one difference: 
instead of $(-1)^{\xi (l) + \pi (l)}$ one has just the factor $-1$ in front of the second term. 
The considerations above imply, that any solution $\wt{A}_{\pi} (\xi)$ is of the form 
$\wt{A}_{\pi} (\xi) = c_{\pi} b (\xi)$, where $c_{\pi}$ is a constant with respect to $\xi$, and 
$b (\xi)$ is some expression of the form $\prod_{i, j} x_{i, j}^{\varepsilon_{i, j} (\xi)}$ with 
$\varepsilon_{i, j} (\xi) = \pm 1$ being some numbers. 
Note, that the fact $|x_{i, j}| \equiv 1$ implies $|b (\xi)| \equiv 1$. 
The constants $c_{\pi}$ can, of course, be chosen 
differently for different values of parameters $\lbrace x_{i, j} \rbrace_{i, j}$, but what is important is that 
$\wt{A}_{\pi} (\xi)$ splits into a product of two factors, one depending only on $\pi$, and the other -- 
having an absolute value $1$ and depending only on $\xi$.     
This allows to establish the unitarity condition in the general case: 
\begin{equation*} 
\sum_{\xi : |\xi| = p} A_{\pi}^{*} (\xi) A_{\pi'} (\xi) = 
c_{\pi}^{*} c_{\pi'} \sum_{\xi : |\xi| = p} a_{\pi}^{*} (\xi) a_{\pi'} (\xi) \, |b (\xi)|^2 = 
2^{N - 2} c_{\pi}^{*} c_{\pi'} \delta_{\pi, \pi'}, 
\end{equation*}
where $|\pi| = |\pi'| = p$, $p \in \mathbb{Z} / 2$. 

\section{Conditions of existence}

It remains to investigate the problem of existence of solutions of the overdetermined linear system 
of equations for $A_{\pi} (\xi)$. 
It suffices to consider only the equations for $A (\xi)$, $|\xi| = 0$.  
For every $\xi$, 
let the notation $\alpha = \lbrace \alpha_{0}, \alpha_{\pm 1} \dots, \alpha_{\pm s} \rbrace$ and 
$\beta = \lbrace \beta_{\pm 1}, \dots, \beta_{\pm q} \rbrace$ be as in the previous section.  
Require that the expression \eqref{A_al_be} for $A (\xi)$ in terms of $\alpha$ and $\beta$ does not depend on the 
choice of numbering of the points of $\alpha$ and $\beta$. 
The independence on the numbering of $\beta$ is seen directly from the formula, so it is necessary to 
focus on $\alpha$. The requirement that 
for any $m$, $1 \leqslant m \leqslant s$, 
it is possible to interchange the numbers of $\alpha_{m}$ and $\alpha_{-m}$, yields a condition: 
\begin{multline*} 
\bigg( 
\prod_{k \in \alpha} \frac{y (k, \alpha_{-m})}{y (k, \alpha_{m})}
\bigg) \, 
\bigg( 
\prod_{k \in \beta} \frac{y (k, \alpha_{m})}{y (k, \alpha_{-m})}
\bigg)
\frac{1}{y^2 (\alpha_{m}, \alpha_{-m})}
= \\ = 
\bigg( 
\prod_{k \in \alpha} \frac{y (k, \alpha_{m})}{y (k, \alpha_{-m})}
\bigg) \, 
\bigg( 
\prod_{k \in \beta} \frac{y (k, \alpha_{-m})}{y (k, \alpha_{m})}
\bigg)
\frac{1}{y^2 (\alpha_{-m}, \alpha_{m})}. 
\end{multline*} 
Rewrite it as follows: 
\begin{multline*} 
y^2 (\alpha_{-m}, \alpha_{m}) \, 
\bigg( 
\prod_{k \in \alpha} y^2 (k, \alpha_{-m})
\bigg) \, 
\bigg( 
\prod_{k \in \beta} y^2 (k, \alpha_{m})
\bigg) 
= \\ =  
y^2 (\alpha_{m}, \alpha_{-m}) \, 
\bigg( 
\prod_{k \in \alpha} y^2 (k, \alpha_{m})
\bigg) \, 
\bigg( 
\prod_{k \in \beta} y^2 (k, \alpha_{-m})
\bigg). 
\end{multline*}
Canceling out the factor $y^2 (\alpha_{-m}, \alpha_{m}) y^2 (\alpha_{m}, \alpha_{-m})$ 
and recalling that $y (i, i) \equiv 1$, one obtains
$Y_{\alpha, \beta} (\alpha_{-m}, \alpha_{m}) = 
Y_{\alpha, \beta} (\alpha_{m}, \alpha_{-m})$, where  
\begin{equation*} 
Y_{\alpha, \beta} (\alpha_{-m}, \alpha_{m}) := \bigg( 
\prod_{\substack{k \in \alpha, \\ k \not = \alpha_{\pm m}}} y^2 (k, \alpha_{-m})
\bigg) \, 
\bigg( 
\prod_{k \in \beta} y^2 (k, \alpha_{m})
\bigg).   
\end{equation*}
In particular, since there exists $\xi$ such that the corresponding $\beta = \emptyset$, one has: 
for all $i, j \in \mathbb{Z} / (N - 1)$, $i \not = j$, 
\begin{equation*} 
\prod_{k : k \not = i, j} y^2 (k, i) = 
\prod_{k : k \not = i, j} y^2 (k, j),  
\end{equation*}
where $k$ varies over $\mathbb{Z} / (N - 1)$. 
Now take the case $\alpha = \lbrace \alpha_0, \alpha_{\pm 1} \rbrace$. Then a product 
$\prod_{k \in \beta}$ is just a product over all $k$, such that $k \not = \alpha_0, \alpha_{\pm 1}$. 
If one writes out the corresponding expressions and then uses the previous formula with 
respect to the pair of points $\lbrace \alpha_1, \alpha_{-1} \rbrace$, one obtains: 
$y^2 (\alpha_{0}, \alpha_{-1}) / y^2 (\alpha_{0}, \alpha_{1}) = 
y^2 (\alpha_{0}, \alpha_{1}) / y^2 (\alpha_{0}, \alpha_{-1})$. 
Hence, for any pairwise non-equal $i$, $j$ and $k$, one has: 
\begin{equation*} 
y^4 (k, i) = y^4 (k, j). 
\end{equation*}
Therefore, each $y (k, j)$, $k \not = j$, is of the form 
$y (k, j) = \lambda (k) z (k, j)$, where $\lambda (k)$ satisfies $|\lambda (k)| = 1$, and 
$z (k, j)$ is a fourth root of unity, $z^4 (k, j) = 1$. 
Recall that by definition $y (i, i) \equiv 1$. 
In particular, if one puts all $z (k, j) = 1$, and substitutes the corresponding $y (k, j)$ into $A (\xi)$, one obtains 
$A (\xi) = (-1)^{s + 1} \wt{A} (\xi)$,  
\begin{equation*} 
\wt{A} (\xi) = \bigg( 
\prod_{m = 1}^{s} \frac{\lambda (\alpha_{m})}{\lambda (\alpha_{-m})} 
\bigg) \, 
\bigg( 
\prod_{m = 1}^{s} \frac{1}{\lambda^2 (\alpha_{m})}
\bigg) \, 
\prod_{k : k \not = \alpha_{0}} \lambda (k) = 
\prod_{k \in \beta} \lambda (k). 
\end{equation*}
In general case, 
taking into account that $(z (k, j))^{2} = \pm 1$, and hence $(z (k, j))^2 = 1 / (z(k, j))^2$, 
the expression for $A (\xi)$ can be transformed into  
\begin{equation*} 
A (\xi) = (-1)^{s + 1} 
\bigg( \prod_{k \in \beta} \lambda (k) \bigg) \, 
\wh{A} (\xi),  
\end{equation*}
where $\wh{A} (\xi)$ is a monomial in variables $z (k, j)$. 
Note, that the first condition (containing $Y_{\alpha, \beta} (\cdot, \cdot)$) on $y (k, j)$ derived 
above, after the substitution 
$y (k, j) = \lambda (k) z (k, j)$, $k \not = j$, yields  
\begin{equation} 
\prod_{k : k \not = i, j} z^2 (k, i) = 
\prod_{k : k \not = i, j} z^2 (k, j).   
\label{cond_pzz}
\end{equation} 
Recall that the second condition $y^4 (k, j) = y^4 (j, k)$ has reduced to  
\begin{equation} 
z^4 (k, j) = 1, \qquad k \not = j, 
\label{cond_z4}
\end{equation}
and, by definition, one has $z (i, i) = 1$. 
Let us derive other conditions. 

Consider again the general formulae \eqref{A_al_be}, \eqref{A_zero_al} for $A (\xi)$. 
Denote by $\alpha_{\pm}$ the sets    
$\alpha_{+} := \lbrace \alpha_1, \alpha_2, \dots, \alpha_s \rbrace$, 
$\alpha_{-} := \lbrace \alpha_{-1}, \alpha_{-2}, \dots, \alpha_{-s} \rbrace$. 
One may pick any two points with opposite indices, say $\alpha_m \in \alpha_{+}$ and 
$\alpha_{-m} \in \alpha_{-}$, and then interchange their locations, i.e. 
put $\alpha_{-m}$ in $\alpha_{+}$ and $\alpha_{m}$ in $\alpha_{-}$. 
The conditions we already have derived ensure that $A (\xi)$ remains invariant.   
Now, for $s \geqslant 2$, it is necessary to require that it is invariant if one switches the indices of any two points 
from the same set, i.e. both from $\alpha_{+}$ or both from $\alpha_{-}$.  
Let us start with $\alpha_{1}$ and $\alpha_{2}$. 
Look at the general formula for $A (\xi)$. Note that the expression in $\lbrace \dots \rbrace$ brackets, and the 
expression for $A (0_{\alpha})$ are already invariant, so it is necessary to focus on the $( \dots )$ 
and $[ \dots ]$ factors. 
Writing out the corresponding equality and canceling out the common factors, one obtains: 
\begin{equation*} 
\frac{y^2 (\alpha_{2}, \alpha_{1}) y^2 (\alpha_{-2}, \alpha_{1})}{y^2 (\alpha_{2}, \alpha_{-2})} = 
\frac{y^2 (\alpha_{1}, \alpha_{2}) y^2 (\alpha_{-2}, \alpha_{2})}{y^2 (\alpha_{1}, \alpha_{-2})}. 
\end{equation*}
Perform the substitutions $y (k, j) = \lambda (k) z (k, j)$, $k \not = j$, $k, j = \alpha_{1}, \alpha_{\pm 2}$, 
and change the notation for indices $\alpha_{1} = p$, $\alpha_{2} = q$, $\alpha_{-2} = r$. This yields 
\begin{equation*} 
\prod_{\substack{i, j \in \lbrace p, q, r \rbrace, \\ i \not = j}} z^2 (i, j) = 1. 
\end{equation*}
This equation has to be true for all pairwise distinct $p$, $q$, $r$. 
Denote $\zeta (i, j) := z^2 (i, j) z^2 (j, i)$, $i \not = j$. Invoking the agreement 
$z (k, k) \equiv 1$, put $\forall k : \zeta (k, k) := 1$. 
Then for \emph{any} $p$, $q$, $r$, not necessarily pairwise distinct, one has a cocycle-type condition:  
\begin{equation} 
\zeta (p, q) \zeta (q, r) \zeta (r, p) = 1. 
\label{zeta_cocycle}
\end{equation}
Note, that $\zeta$ is symmetric, $\zeta (i, j) = \zeta (j, i)$, and  normalized on $1$. 
Hence, $\zeta (i, j)$ in \eqref{zeta_cocycle} is of the form $\zeta (i, j) = \varphi (i) \varphi (j)$, where 
$\varphi : \mathbb{Z} / (N - 1) \to \mathbb{C}$ is some function, such that $\varphi^2 (i) \equiv 1$, 
i.e. $\varphi (i) = \pm 1$. 
Observe, that since $N - 1$ is odd, for any $s \in \mathbb{Z} / (N - 1)$ we have: 
$[s]_2 = 0$ iff $[-s]_2 = 1$ 
(recall, that $[\cdot]_{2} : \mathbb{Z} / (N - 1) \twoheadrightarrow \mathbb{Z} / 2$ denotes the 
canonical epimorphism). 
If one assigns the values of $z^2 (k, k + l)$, for all $k$ and, say, only for all $l$, $[l]_2 = 1$, 
then one can extend the function $z^2 (\cdot, \cdot)$ to all points  
according to $z^2 (k, k - l) = \varphi (k) \varphi (k - l) / z^2 (k - l, (k - l) + l)$ 
($l$ satisfies $[l]_2 = 1$), 
$z^2 (k, k) = 1$. Since $z^2 (i, j)$ can only be $\pm 1$, and $\varphi (i)$ can only be $\pm 1$, we have: 
$\varphi (i) z^2 (i, j) = \varphi (j) z^2 (j, i)$, for all $i, j$, $i \not = j$. 
Recalling, that $y (k, l) = \lambda (k) z (k, l)$, $k \not = l$, 
and that there are no any conditions on $\lambda (k)$ so far except $|\lambda (k)| = 1$, 
we see, that the factor $\varphi (\cdot)$ is unimportant and can be incorporated into $\lambda (\cdot)$. 
Therefore, without loss of generality, $y (k, l) = \lambda (k) z (k, l)$, $k \not = l$, and 
for all $i$, $j$: 
\begin{equation} 
z^2 (i, j) = z^2 (j, i).  
\label{cond_z2_perm}
\end{equation}
The latter assumption allows to rewrite $\wh{A} (\xi)$ in the form:  
\begin{multline} 
\wh{A} \bigg( 
\begin{matrix} 
\text{$0$ at $\alpha$,}\\
\text{$1$ at $\beta$}
\end{matrix}
\bigg) := 
\bigg( 
\prod_{k \in \alpha} \frac{
\prod_{l \in \alpha_{-}} z (k, l)
}{ \prod_{l \in \alpha_{+}} z (k, l)}
\bigg) \, 
\bigg( 
\prod_{k \in \beta} \frac{
\prod_{l \in \alpha_{+}} z (k, l)}{
\prod_{l \in \alpha_{-}} z (k, l)}
\bigg)
\times \\ \times 
\bigg( \prod_{\substack{i, j \in \alpha \backslash \lbrace \alpha_{0} \rbrace, \\ i < j}}
z^2 (i, j) \bigg) 
\prod_{k = 0}^{N - 2} z (k, \alpha_0). 
\label{Ahat_al_be}
\end{multline}
It is now clear, that we have an expression for $A (\xi)$, which does not feel the choice of 
numbering of points in $\alpha_{+}$ and $\alpha_{-}$.  
Recalling the fact, that for any $m$ one may interchange the roles of $\alpha_{m}$ and $\alpha_{-m}$, 
we conclude, that $A (\xi)$ does not depend on the choice of partitioning of 
$\alpha \backslash \lbrace \alpha_{0} \rbrace$ into $\alpha_{+}$ and $\alpha_{-}$. 
It remains to investigate the requirement of no dependence on the choice of the marked point 
$\alpha_{0} \in \alpha$. 

Without loss of generality, look at $\alpha_{0}$ and $\alpha_{1}$. 
One obtains: 
\begin{multline*} 
\bigg( \prod_{\substack{j \in \alpha, \\ j \not = \alpha_0, \alpha_1}} 
z^2 (\alpha_1, j) \bigg) \, 
Z_{\alpha, \beta} (\alpha_{-1}, \alpha_1) \, 
\prod_{k} z (k, \alpha_0) 
= \\ =   
\bigg( \prod_{\substack{j \in \alpha, \\ j \not = \alpha_0, \alpha_1}} 
z^2 (\alpha_0, j) \bigg) \, 
Z_{\alpha, \beta} (\alpha_{-1}, \alpha_{0}) \, 
\prod_{k} z (k, \alpha_1).   
\end{multline*}
where
\begin{equation*}
Z_{\alpha, \beta} (i, j) := 
\bigg( \prod_{k \in \alpha} 
\frac{z (k, i)}{ z (k, j)} 
\bigg) \, 
\bigg( \prod_{k \in \beta} 
\frac{z (k, j)}{z (k, i)} \bigg),  
\end{equation*}
$i, j \in \mathbb{Z} / (N - 1)$. Canceling out the common factors and 
getting rid of denominators, one arrives at 
\begin{equation*} 
z^2 (\alpha_1, \alpha_0) 
\prod_{\substack{i \in \alpha, \\ i \not = \alpha_0, \alpha_1}} z^2 (\alpha_1, i) z^2 (i, \alpha_0) = 
z^2 (\alpha_0, \alpha_1) 
\prod_{\substack{i \in \alpha, \\ i \not = \alpha_0, \alpha_1}} z^2 (\alpha_0, i) z^2 (i, \alpha_1). 
\end{equation*}
But the latter equality is implied directly by the assumption $z^2 (i, j) \equiv z^2 (j, i)$. 
Therefore, no new conditions on $z (i, j)$ emerge. 

We have a well-defined expression for $A (\xi)$, i.e. it does not depend on the choice of 
numbering of points in $\alpha := \lbrace i \, | \, \xi (i) = 0 \rbrace$ and 
$\beta := \lbrace j \, | \, \xi (j) = 1 \rbrace$. 
Still it doesn't prove that $A (\xi)$ satisfies the whole system equations. 
It is necessary to verify that the substitution of $A (\xi)$ into the corresponding system indeed 
turns each of the equations into an identity. This is equivalent to verifying  
\begin{equation*} 
\bigg( \prod_{\substack{k \in \alpha, \\ k \not = l}} z (k, l) \bigg) 
\wh{A} \bigg( 
\begin{matrix} 
\text{$1$ at $\alpha$}, \\ \text{$0$ at $\beta$}
\end{matrix}
\bigg) = 
\bigg( \prod_{\substack{k \in \beta, \\ k \not = l}} z (k, l) \bigg) 
\wh{A} \bigg( 
\begin{matrix}
\text{$1$ at $\beta \Delta \lbrace l \rbrace$}, \\ 
\text{$0$ at $\alpha \Delta \lbrace l \rbrace$} 
\end{matrix}
\bigg), 
\end{equation*}
for all $l \in \mathbb{Z} / (N - 1)$. 
With the notation as above, it suffices to consider just two cases: 
$l = \alpha_{0}$ and $l = \beta_{q}$. 

Specialize first to the case $l = \alpha_0$. 
The sets $\alpha' := \beta \Delta \lbrace l \rbrace$ and 
$\beta' := \alpha \Delta \lbrace l \rbrace$ are of the form 
$\alpha' = \beta \cup \lbrace \alpha_0 \rbrace$, $\beta' = \alpha \backslash \lbrace \alpha_{0} \rbrace$. 
As above, fix some numbering for $\alpha$ and $\beta$, $\# \alpha = 2 s + 1$, $\# \beta = 2 q$. 
It is natural to equip $\alpha'$ and $\beta'$ with the following numbering: 
$\alpha_{0}' := \alpha_{0}$, $\alpha_{j}' := \beta_{j}$, $j = \pm 1, \pm 2, \dots, \pm q$; 
$\beta_{i}' := \alpha_{i}$, $i = \pm 1, \pm 2, \dots, \pm s$. 
Substitute the corresponding expressions \eqref{Ahat_al_be} for $\wh{A} (\cdot)$ into the last equation and 
simplify the result as follows. 
First of all, get rid of all denominators and write the equation as an equality between two   
products of products of the form $\prod_{i, j} z^t (i, j)$, $t = 1$ or $2$. After that, split each of the 
products $\prod_{i, j} z^{t} (i, j)$ into several factors of the form $\prod_{i \in I, j \in J} z^t (i, j)$, 
where $I$ and $J$ are one of the following sets: $\alpha_{+}$, $\alpha_{-}$, $\beta_{+}$, $\beta_{-}$, or 
a one-point set $\lbrace \alpha_{0} \rbrace$. 
Simplify the expressions using the assumptions $z^2 (i, j) = z^2 (j, i)$ and $z^4 (i, j) = 1$, and 
invoking the convention $z (i, i) = 1$.  
Cancel out the common factors in the left and right-hand sides of the equation. 
Finally, dividing the left over the right-hand side, one arrives at: 
\begin{equation*} 
\bigg( 
\prod_{\substack{k, j = 0, \\ k < j}}^{N - 2} 
\frac{z (k, j)}{z (j, k)}
\bigg) \, 
\bigg[ 
\prod_{\substack{k \in \alpha_{-} \cup \beta_{-}, \\ 
j \in \alpha_{+} \cup \beta_{+} \cup \lbrace \alpha_{0} \rbrace}}
z^2 (k, j)
\bigg] = 1. 
\end{equation*} 
Note, that $z (k, j) / z (j, k)$ is symmetric with respect to the permutation of $k$ and $j$. 
Denote the expression in the square brackets by $\Phi (\alpha_{-} \cup \beta_{-})$. 
Since $z^2 (i, j) = z^2 (j, i)$ and $z^2 (i, i) = z^2 (j, j) = 1$, one has 
$\prod_{k} z^2 (k, i) = \prod_{k} z^2 (k, j)$, where $k$ varies over the entire $\mathbb{Z} / (N - 1)$. 
Hence, there exists $g \in \lbrace \pm 1 \rbrace$, such that for all $i$ one has $\prod_{k} z^2 (k, i) = g$. 
In particular, for all $k \in \alpha_{-} \cup \beta_{-}$ this allows to derive 
$z^2 (k, \alpha_{0}) = z^2 (\alpha_{0}, k) = g
\prod_{m \in (\alpha_{-} \cup \beta_{-}) \cup (\alpha_{+} \cup \beta_{+})} z^2 (m, k)$. 
Expressing $z^2 (k, \alpha_{0})$ this way in $\Phi (\alpha_{-} \cup \beta_{-})$, regrouping the factors, 
and taking into account that $\# (\alpha_{-} \cup \beta_{-}) = (N - 2)/ 2$ is odd, one derives  
\begin{equation*} 
\Phi (\alpha_{-} \cup \beta_{-}) = 
g \, \bigg( \prod_{\substack{k \in \alpha_{-} \cup \beta_{-}, \\ 
j \in \alpha_{+} \cup \beta_{+}}} z^4 (k, j) \bigg) \, 
\prod_{\substack{k, j \in \alpha_{-} \cup \beta_{-}, \\ k < j }} z^4 (k, j) 
= g. 
\end{equation*} 
Therefore, one has a condition: $\prod_{k, j : k < j} z (k, j)/ z (j, k) = g$. 
Squaring the left and the right-hand sides, taking a product over all $k$, 
and using the assumptions about $z^2 (i, j)$, one derives: 
$\prod_{k, j = 0}^{N - 2} z^2 (k, j) = g^2 = 1$. 
One the other hand, the latter product is just 
$\prod_{k = 0}^{N - 2} (\prod_{j = 0}^{N - 2} z^2 (k, j)) = g^{N -1} = g$. 
Therefore, $g = 1$. 
Recall that $z (k, j)/ z (j, k)$ is symmetric with respect to the permutation of $j$ and $k$.  
Hence, one may take \emph{any} relation $Q$ on 
$\mathbb{Z} / (N -1)$, such that $\forall k, j, \, j \not = k : (k, j) \in Q 
\Leftrightarrow (j, k) \not \in Q$, and rewrite the product 
in the $(\dots)$ factor 
in the condition above as $\prod_{(k, j) \in Q} z(k, j)/ z (j, k)$.  
In particular, it is possible to take $Q$ being formed by all ordered pairs $(k, j)$, 
such that $[j - k]_2 = 1$  
(recall, that since $N - 1$ is odd, $[j - k]_2 = 1$ iff $[k - j]_2 = 0$). 
Then the condition on $z (k, j)$ acquires the form: 
\begin{equation} 
\prod_{\substack{k, m \in \mathbb{Z} / (N - 1), \\ [j - k]_2 = 1}} z (k, j) = 
\prod_{\substack{k, j \in \mathbb{Z} / (N - 1), \\ [j - k]_2 = 1}} z (j, k). 
\label{cond_pz_pz}
\end{equation}  
But this condition implies that $\prod_{j : j \not = k} z^2 (k, j)$ equals to 
\begin{equation*} 
\prod_{m : [m]_2 = 1} z^2 (k, k + m) z^2 (k, k - m) = 
\prod_{m : [m]_2 = 1} z^4 (k, k + m) = 1,  
\end{equation*}
i.e. the condition \eqref{cond_pzz} is implied by \eqref{cond_pz_pz}, \eqref{cond_z2_perm}, \eqref{cond_z4}. 

Now consider the case $l = \beta_{q}$. Essentially it is investigated as the case above. 
For this $l$ one has $\alpha'' := \beta \Delta \lbrace l \rbrace = 
\beta \backslash \lbrace \beta_q \rbrace$, 
$\beta'' := \alpha \Delta \lbrace l \rbrace = \alpha \cup \lbrace \beta_q \rbrace$. 
Hence $\# \alpha'' = 2 q - 1$, and $\# \beta'' = 2 s + 2$. 
It is natural to put $\alpha_{0}'' = \beta_{-q}$, $\alpha_{+}'' = \beta_{+} \backslash \lbrace \beta_{q} \rbrace$, 
$\alpha_{-}'' = \beta_{-} \backslash \lbrace \beta_{-q} \rbrace$, 
$\beta_{+}'' = \alpha_{+} \cup \lbrace \alpha_{0} \rbrace$, and 
$\beta_{-}'' = \alpha_{-} \cup \lbrace \beta_{q} \rbrace$. 
The following equality is required: 
\begin{equation*} 
\bigg( \prod_{\substack{k \in \alpha, \\ k \not = \beta_{q}}} z (k, \beta_q{}) \bigg) 
\wh{A} \bigg( 
\begin{matrix} 
\text{$0$ at $\alpha$}, \\ \text{$1$ at $\beta$}
\end{matrix}
\bigg) 
= 
\bigg( \prod_{\substack{k \in \beta, \\ k \not = \beta_{q}}} z (k, \beta_q{}) \bigg) 
\wh{A} \bigg( 
\begin{matrix} 
\text{$0$ at $\alpha''$}, \\ \text{$1$ at $\beta''$}
\end{matrix}
\bigg). 
\end{equation*} 
Substitute the corresponding expressions for 
$\wh{A} (\cdot)$ (see \eqref{Ahat_al_be}) into this equality and then transform it as follows. 
Get rid of the denominators, and then rearrange the factors in the left and right-hand sides 
so that each becomes a product of products of the form $\prod_{i \in I, j \in J} z^t (i, j)$, 
where $t = 1$ or $2$, and  
$I$ and $J$ are one of the following sets: $\alpha_{+}$, $\alpha_{-}$, 
$\beta_{+} \backslash \lbrace \beta_{q} \rbrace$, 
$\beta_{-} \backslash \lbrace \beta_{-q} \rbrace$, 
$\lbrace \alpha_{0} \rbrace$, $\lbrace \beta_{q} \rbrace$, or $\lbrace \beta_{-q} \rbrace$.  
Simplify the expressions using $z (i, i) = 1$, $z^2 (i, j) = z^2 (j, i)$ and $z^4 (i, j) = 1$. 
Finally, divide the left-hand side over the right-hand side, and simplify the result again. 
A brute force calculation yields: 
\begin{equation*} 
\bigg( 
\prod_{\substack{k, j = 0, \\ k < j}}^{N - 2} 
\frac{z (k, j)}{z (j, k)}
\bigg) \, 
\Phi \big( 
( \beta_{-} \backslash \lbrace \beta_{-q} \rbrace ) \cup 
( \alpha_{-} \cup \lbrace \beta_{q} \rbrace )
\big) = 1. 
\end{equation*} 
But the $\Phi$ factor in the left-hand side is itself equal to $1$, so one arrives just 
at the condition \eqref{cond_pz_pz} already derived above. 

Finally, it remains to recall that $x_{i, j} = \lambda ((i + j)/ 2) z ((i + j)/ 2, j)$, $i \not = j$, 
and that $\forall i, j, \, i \not = j: x_{i, j} = x_{j, i}$. 
This yields $z ((i + j)/ 2, i) = z ((i + j)/ 2, j)$, 
or, what is equivalent, $\forall k, m : z (k, k + m) = z (k, k - m)$. 
Rewriting the condition \eqref{cond_pz_pz} on $z$ in the form 
$\prod_{k} \prod_{m : [m]_2 = 1} z (k, k + m) = 
\prod_{k} \prod_{m : [m]_2 = 1} z (k, k - m)$, 
one can see, that it is satisfied.  
Next, since the square $z^2 (i, j)$ is symmetric with respect to the permutation of $i$ and $j$, 
one can always represent $z (k, j)$ in the form $z (k, j) = \varkappa (2 k - j, j) \mu (k, j)$, where 
$\forall k, j : \mu (k, j) = \mu (j, k)$, and $\forall i, j : \varkappa^2 (i, j) = 1$. 
One has $\mu^4 (k, j) = 1$ and $\mu^2 (k, k + m) = \mu^2 (k, k - m)$. 
It is always possible to adjust $\mu (\cdot, \cdot)$ and $\varkappa (\cdot, \cdot)$ so that 
$\forall k, m : \mu (k, k + m) = \mu (k, k - m)$.   
The formula for $x_{i, j}$, $i \not = j$,  now becomes 
\begin{equation} 
x_{i, j} = \varkappa (i, j) 
\lambda \Big( \frac{i + j}{2} \Big) 
\mu \Big( \frac{i + j}{2}, j \Big). 
\label{kappa_lambda_mu}
\end{equation}
Since $x_{i, j} = x_{j, i}$, there is a condition $\forall i, j : \varkappa (i, j) = \varkappa (j, i)$. 
Note, that if one takes any $\mu$ and $\varkappa$ with the mentioned properties, and 
puts $z (k, j) = \varkappa (2 k - j, j) \mu (k, j)$, then one has automatically $\forall k, m : 
z (k, k + m) = z (k, k - m)$, together with all other required properties. 

The functions $\mu$ and $\varkappa$ in \eqref{kappa_lambda_mu} 
can be interpreted as follows. Consider a unit circle in $\mathbb{C}$ centered at zero, 
and mark on it the points $e^{i 2 \pi m / (N - 1)}$, $m = 0, 1, \dots, N - 2$. 
Identify naturally these points with the elements of $\mathbb{Z} / (N - 1)$. 
Look at all chords connecting the marked points. To define a function $\mu$ is the same as to 
define a function on all these chords with values in the roots of unity of degree $4$. 
Whenever two chords have the same length and share a common vertex, the corresponding values of 
the function must coincide. Note, that if $N - 1$ is prime, then  
this implies that the value $\mu (i, j)$ is determined by $j - i$.   
The definition of $\varkappa$ in \eqref{kappa_lambda_mu} is equivalent to 
assigning the numbers $\pm 1$ to the chords in an arbitrary manner.  

This completes the investigation of the conditions of solvability of the system for $A_{\pi} (\xi)$. 
Since these conditions can be satisfied, one is able to construct new examples of non-bicolorable configurations of rays. 
Let us summarize the results. 
Recall, that the described construction involves a set $V := (\mathbb{Z} / (N - 1)) \sqcup \lbrace \ast \rbrace$, 
where $\ast$ is a formal symbol. To every $v \in V$, a collection of rays 
$\Psi_{\sigma}^{v}$, $\sigma \in L (\lbrace v \rbrace)$, in $(\mathbb{C}^2)^{\otimes (N - 1)}$ is assigned, and  
there is also one more collection of rays $F_{\pi}$, $\pi \in L (V)$, defined in terms of $A_{\pi} (\xi)$. 
Invoking the notation \eqref{index_set_LU} and that $p_0 = 1$, $p_1 = 0$, one has  
$L (\lbrace v \rbrace) = \lbrace \phi : V \to \mathbb{Z} / 2 \, | \, \phi (v) = 0 \rbrace$, and 
$L (V) = \lbrace \phi : V \to \mathbb{Z} / 2 \, | \, \sum_{z \in V} \phi (z) = 1 \rbrace$.

\begin{thm}
Let $N \in \mathbb{N}$ be a positive integer divisible by $4$. 
Let $\lambda : \mathbb{Z} / (N - 1) \to \mathbb{C}$ and 
$\mu, \varkappa : (\mathbb{Z} / (N - 1))^2 \to \mathbb{C}$ be any functions 
having the values on a unit circle, in roots of unity of degree $4$, and in roots of unity of degree $2$, respectively. 
Assume that both  $\mu$ and $\varkappa$ are symmetric with respect to the permutation of their two arguments, and that 
$\forall j, k : \mu (k, k + j) = \mu (k, k - j)$. 
Let $\lbrace \varphi [k]_{\alpha} \rbrace_{\alpha \in \mathbb{Z} / 2}$, $k \in \mathbb{Z} / (N - 1)$, be any 
family of orthonormal bases in $\mathbb{C}^2$. 
For every $i \not = j$, put 
\begin{equation*} 
x_{i, j} := \varkappa (i, j) 
\lambda \Big( \frac{i + j}{2} \Big) 
\mu \Big( \frac{i + j}{2}, j \Big).
\end{equation*} 
Define the rays 
$\Psi_{\rho}^{v}$, $\rho \in L (\lbrace v \rbrace)$, 
$v \in V := (\mathbb{Z} / (N - 1)) \sqcup \lbrace \ast \rbrace$, according to  
\eqref{ray_Psi_star}, \eqref{ray_Psi_l}, using \eqref{matrix_u}, \eqref{special_u}. 
Define the rays $F_{\pi}$, $\pi \in L (V)$, according to 
\eqref{ray_F}, using \eqref{A_pi_xi_null}, \eqref{A_xi}, \eqref{A_al_be}, \eqref{A_zero_al}. 
Claim, that the finite configuration in $(\mathbb{C}^{2})^{\otimes (N - 1)}$ formed by these rays is non-bicolorable.  
\qed 
\end{thm}

\section{Discussion} 

In the present paper a new infinite family of examples of non-bicolorable configurations of rays has been described. 
More precisely, it is better to view it as a family of families of examples: there is a parameter 
$N = 4 n$, $n \in \mathbb{N}$, indexing the families, and for each $N$ there are several complex-valued parameters 
(their number depends on $N$), which index the configurations. 
If one puts $N = 4$ and all $x_{i, j} = 1$, then one recovers just the configuration of rays 
described in \cite{Mermin, KernaghanPeres}, but in completely different notation.  

Observe, that the projective lines $\Psi_{\rho}^{\ast}$, $\rho \in L (\lbrace \ast \rbrace)$, 
don't depend on $x_{i, j}$. 
The relation $\not \perp$ depends on $x_{i, j}$, but at the same time it 
doesn't change if one varies the continuous part of $x_{i, j}$, i.e. the function $\lambda$. 
Hence, for a given $N$, the corresponding configurations we have can be 
viewed as deformations of each other in the sense of the definition given above. 
It is natural to mark the configuration corresponding to all $\lambda (k) = 1$, and view the others as its 
deformations. 

We have a notion of deformation which connects two configurations. 
Its definition implies that either both configurations are bicolorable, or both are 
non-bicolorable. It is possible to extend this definition so that to capture the transition 
\emph{bicolorable} $\longrightarrow$ \emph{non-bicolorable}, and vice versa, but this 
stays beyond the framework of the present paper. 
It is, of course, necessary to replace the requirement that the bijection in the definition of deformation 
respects the $\not \perp$ relation by something else. 
Intuitively, this should be a requirement that the bijection respects the ``template'' of the relation, 
but not the relation itself. For example, in the definition of the relation $R$ there are four 
parameters $p_0, p_1, p_2, p_3$. 
Varying the values of the parameters, one obtains different $R$, but of a similar form. 
In order to formulate a consistent generalization, it appears natural to consider \emph{saturated} 
configurations, i.e. such ones, for which any subset of pairwise orthogonal rays can be embedded in a subset 
of $d$ pairwise orthogonal rays, $d$ -- the dimension of space $\mathcal{H}$. 
So it is necessary to construct a saturation for the new examples first. 

The other direction of possible generalizations is to increase the number of colors. 
Note, that it is related to the interpretation of Kochen-Specker theorem in 
terms of generalized valuations in 
\cite{IshamButterfield, ButterfieldIsham, HamiltonIshamButterfield, ButterfieldIsham1}.
A reasonable definition of \emph{non-colorable} configuration for several colors 
will, apparently, require an introduction of some filtration on 
$\mathbb{P} (\mathcal{H}) \times \mathbb{P} (\mathcal{H})$, which is 
meant to replace the $\not \perp$ relation. 
This filtration might depend on a particular physical problem. 
The set of colors, present in the definition, would also require some 
additional structure, similar to the one of an orthoalgebra.

Intuitively, a saturated configuration has to possess some symmetry. 
For the configuration corresponding to $N = 4$ and all $x_{i, j} = 1$, a finite saturation has been 
constructed in \cite{RuugeFVO}, motivated by \cite{Ruuge}. 
Its symmetry turns out to be described by a non-trivial Abelian 
extension $\mathcal{G}$ of $(\mathbb{Z} / 2)^{6}$ over $GL (4, \mathbb{F}_2)$ (the general linear 
group of $4 \times 4$ matrices over a field with two elements). The number $6$ stems from some combinatorics  
and should be viewed as a binomial coefficient $C_{4}^{2}$. Hence, one looks at a short exact sequence in the 
category of groups 
\begin{equation}  
0 \to (\mathbb{Z} / 2)^{6} \to \mathcal{G} \to GL (4, \mathbb{F}_2) \to 0. 
\label{exact_seq}
\end{equation}
The point of view described in that paper 
appears to be quite general. 
It might be possible to construct new 
finite saturated non-bicolorable configurations by way of considering 
in analogy with \eqref{exact_seq}
the extensions $0 \to A \to \wt{G} \to G \to 0$, for $G$ and $A$ being some other finite groups, 
and $A$, say, elementary Abelian. 
The condition of non-bicolorability then yields simply a condition on the corresponding 
cohomology class $\alpha \in H^2 (G, A)$. 

\section*{Acknowledgements} 

The author would like to thank professor F. Van Oystaeyen for interesting discussions and 
the anonymous referee for important remarks. 
The present work has been financially supported by the Liegrits programme of the European Science Foundation.

\end{document}